\documentclass{article}

\usepackage[utf8]{inputenc}
\usepackage[T1]{fontenc}

\usepackage[english]{babel}

\usepackage[letterpaper,top=2cm,bottom=2cm,left=3cm,right=3cm,marginparwidth=1.75cm]{geometry}

\usepackage{amsmath,amssymb,amsfonts}
\usepackage{amsthm}
\newtheorem{definition}{Definition}

\usepackage{chngcntr}
\counterwithout*{figure}{section}

\usepackage[colorlinks=true, allcolors=blue]{hyperref}
\usepackage{graphicx}
\usepackage{tikz}
\usetikzlibrary{shapes}
\usetikzlibrary{arrows.meta}
\usepackage{subcaption}
\usepackage{booktabs}
\usepackage{multirow}
\usepackage{threeparttable}
\usepackage{longtable}
\usepackage{tabularx}
\usepackage{ragged2e}
\newcolumntype{L}{>{\RaggedRight\arraybackslash}X}
\usepackage{xltabular}
\usepackage{enumitem}
\usepackage[numbers,sort&compress]{natbib}
\usepackage{csquotes}
\newlist{tabitem}{itemize}{1}
\setlist[tabitem]{
    leftmargin=*,
    nosep,
    topsep=0pt,
    partopsep=0pt,
    parsep=0pt,
    itemsep=2pt
}
\title{People, Places \& Things:\\ Network topology \& motifs of R\&D missions}

\author{Henry C.W. Price\textsuperscript{1,2,3,4\S}, Martin Ho\textsuperscript{1,2\S*}, 
Tim S. Evans\textsuperscript{3,4}, 
Eoin O'Sullivan\textsuperscript{1,2}}

\begin{document}
\maketitle

\begin{center}

	\textbf{1} Centre for Science Technology \& Innovation Policy, University of Cambridge, Cambridge CB3 0HU, United Kingdom
	\\
	\textbf{2} Department of Engineering,  University of Cambridge, Cambridge CB3 0HU, United Kingdom
	\\
	\textbf{3} Centre for Complexity Science, Imperial College London, London SW7 2AZ, United Kingdom
	\\
	\textbf{4} Theoretical Physics Group, Department of Physics, Imperial College London, London SW7 2AZ, United Kingdom
	\\[0.5\baselineskip]
	
	\S These authors contributed equally to this work.
	\\
    * Corresponding author: \texttt{wtmh3@eng.cam.ac.uk}

\end{center}
\begin{abstract}
\noindent 
{Challenge-led R\&D programs increasingly assemble heterogeneous people, organizations, funders, projects, and technical outputs around defined missions. Yet program evaluation often describes these systems through project lists, output counts, or retrospective case narratives. This article develops a typed network framework for representing R\&D program architecture directly. We model programs as networks of people, places, and things: researchers, program directors, institutions, funders, publications, patents, projects, and citations. Applied to ARPA-E project impact sheets from the agency's first decade, the framework reconstructs 23 program-induced networks and an agency-level composed network. We show that R\&D programs have an analysable topology: a typed arrangement of people, institutions, funders, projects, publications, patents, and citations that can be reconstructed, compared, and monitored. The analysis shows that programs can be compared by their local structural patterns, that cross-program overlap is concentrated more in recurring institutions than in individual researchers, and that program fingerprints differ across thematic areas. The article contributes to network science by extending topological analysis to R\&D program systems, a class of governed, typed, and output-generating networks that has not been systematically represented in existing innovation-network work.}

\end{abstract}

\section{Introduction}
\label{ch:intro}

{R\&D programs are often designed as structured interventions but evaluated through disaggregated lists of activities and outputs. Program managers assemble performers, institutions, funders, projects, and technical outputs around a mission, yet this architecture is often flattened into separate counts of publications, patents, awards, firms, or follow-on funding. The result is a mismatch between how programs are designed and how they are studied \cite{RN182,RN276,RN280,RN274,RN285}.}

{This paper addresses that mismatch by representing R\&D programs as typed networks of people, places, and things. The purpose is not to reduce programs to a single performance indicator, nor to infer causal impact from network position, but to make program architecture visible: who and what is connected, through which relation, at what level of the program hierarchy, and in what recurring local structures.}

{Network science has shown that different complex systems exhibit characteristic topologies. Shipping systems, supply chains, gene regulation networks, communication infrastructures, the Internet, and social networks each have recurring structural forms that shape flow, robustness, coordination, and diffusion. R\&D programs also depend on flow, coordination, and diffusion, yet their topology has rarely been specified. We extend network science by treating R\&D programs as a distinct empirical network class: governed, typed, directed, program-induced, organizationally scaffolded, output-generating networks whose topology reflects both deliberate program design and emergent scientific and technological linkage.}

{For R\&D management, this matters because program managers do not only fund projects; they assemble capabilities. They choose performers, structure portfolios, connect universities and firms, stage outputs, and make decisions about whether a program should diversify across many teams or concentrate around a few anchors. For policy, the same representation supports structural monitoring of mission-oriented agencies as they proliferate across government, philanthropy, and industry \cite{RN3065,RN3077,RN36,RN45,RN160,RN3041,RN2859,RN3118}. Table \ref{tab:why_networks} outlines situations in which network science is an especially suitable analytical device.}

\begin{table}
\caption{When is network science an appropriate analytical instrument for R\&D program?}
\centering
\label{tab:why_networks}
\small
\begin{threeparttable}
    \begin{tabular}{p{0.3\textwidth}>{\RaggedRight}p{0.3\textwidth}p{0.3\textwidth}}
        \toprule
        \textbf{Relational properties only recoverable via networks} & \textbf{R\&D programs properties that await quantification} & \textbf{Non-R\&D scenarios where network models exist} \\ 
        \toprule
        \textbf{Relationships:} Entities interact rather than simply expressing individual attributes & Team science that requires collaboration across disciplines and TRLs (e.g. basic research, testbeds, demonstrators, manufacturing) \cite{RN19} & Friendship networks and key influencers, financial transaction networks \\ 
        \midrule
        \textbf{Higher-order structure:} Paths, communities, motifs, connectivity patterns that explain overall system behaviors & Technological trajectories, emergence, and convergence within an innovation system \cite{RN367} & Bipartite graphs in recommender systems, fraud detection networks  \\ 
        \midrule
        \textbf{Naturally relational and non-Euclidean data structure:} Non-rigid structures that require flexible representation & Innovation contains many-to-many multivariate connections among many different types of collaborators, infrastructure, funding sources, outputs, and outcomes \cite{RN276}& Molecular structure, transport networks, supply chains \\ 
        \midrule
        \textbf{Influence, diffusion \& flow of information:} Knowledge relies on spillover and absorption to exert impact & R\&D is a dynamic process with diffusion, translation, and dependencies occurring between different innovation lifecycles \cite{RN285} & Information diffusion (e.g. viral tweets), epidemic spread, energy flow \\ 
        \bottomrule
    \end{tabular}
\begin{tablenotes}\footnotesize
 \item Authors' own analysis. Network representation should be prioritized when connections are more important than the individual parts of R\&D. Network should not be used when data is lacking, weak, noisy, or independently distributed.
\end{tablenotes}
\end{threeparttable}
\end{table}

{The crux of this study lies in examining the various \emph{interactions} induced by R\&D programs: including the people (performers \& agencies) involved, the places (institutions, infrastructure) where innovation occurs, and the things (innovations) that are produced. When it comes to studying innovation networks, scholars have conventionally inspected citations or topic co-occurrences \cite{RN1374,RN997,RN1247,RN1427,RN1625,RN1746} from the top down. Here we instead model the social, organizational, and technological layers of the same set of R\&D programs as a single typed network and analyze their local structural patterns from the bottom up. Our study asks:}

\begin{itemize}
    \item Can the structural architecture of R\&D programs be reconstructed using networks?
    \item {Within these networks, can the structural configuration and communication patterns among {researchers, funding organizations, and knowledge outputs} in R\&D programs be objectively and systematically characterized?}
\end{itemize}

{To answer these questions, we define the R\&D program as the primary network object. Each program is represented as a typed, directed, program-induced subgraph of a wider agency network, and each subgraph can then be inspected internally, compared externally, and composed into an agency-level innovation system.}

{The central contribution is therefore representational and diagnostic. We formalize typed, directed multi-graphs that unify funding relations, collaboration, institutional affiliations, output production, ownership, and citation-linked structures. This allows us to move beyond aggregate indicators and recover the configurations disclosed around project outputs. Crucially, by representing each program as an induced subgraph of a shared global network, we can detect structural overlaps between programs that are invisible when programs are analyzed in isolation. We show how program-level ego networks, induced subgraphs, and typed local patterns can (i) visualize the social and institutional architecture of R\&D projects, (ii) trace citation-linked and funder-linked structures around those projects, and (iii) quantify patterns of cross-program overlap and institutional bridges that remain invisible in standard project-level monitoring.}

{Our central contribution to network science is to treat R\&D programs not only as an application domain for existing network measures, but as network objects in their own right. We specify their topology, define their boundaries, identify their local structural primitives, and show how program-level networks compose into an agency-level innovation system.}

{The purpose of the framework is observational and diagnostic rather than hypothesis-testing: it provides a way to observe and compare program architectures that are otherwise distributed across heterogeneous records.}

The remainder of this article shifts the interrogation of R\&D programs from qualitative system descriptions and indirect empirical proxies toward quantitative architectural measurement. Much existing work -- whether descriptive in orientation or directed toward causal explanation -- infers program structure through partial indicators such as budgets, outputs, organizational characteristics, or retrospective case evidence. These approaches are informative, but they do not directly observe the relational architecture through which R\&D programs are organized and governed. This article addresses that measurement gap by developing a mathematical language for representing and analyzing R\&D program networks.  Section \ref{ch:methods} sets up the mathematical language to measure R\&D program networks. Section \ref{ch:data} introduces the dataset and workflow to create and analyze networks. Section \ref{ch:results_visuals} applies network-specific analyses to elucidate program management practices. Section \ref{ch:discussion} discusses implications for mission-oriented R\&D. Section \ref{ch:conclusion} concludes.

\section{Methods}
\label{ch:methods}

\begin{figure}[h]
    \centering
    \includegraphics[width=0.76\paperwidth]{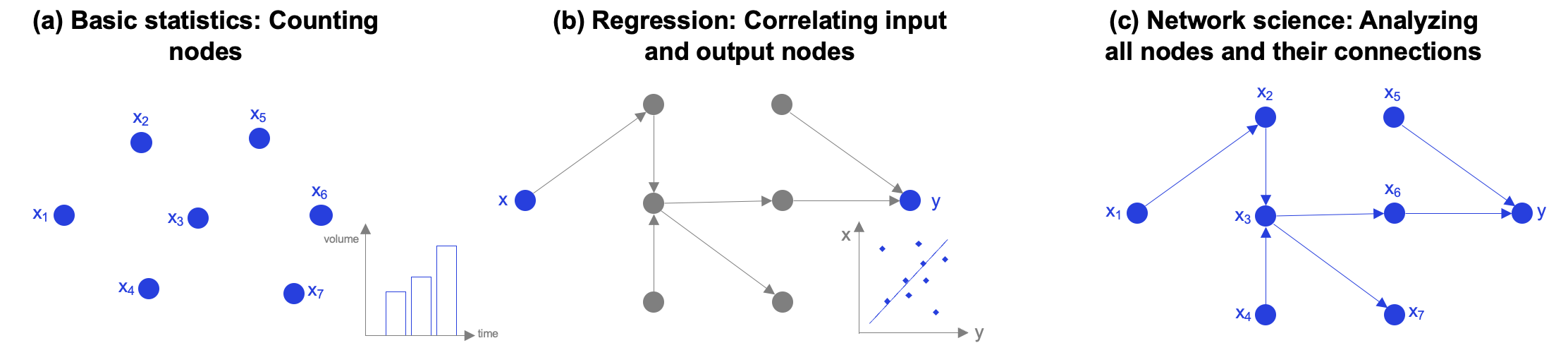}
    \caption{(a) Simple tally of raw/curated innovation data (e.g. citations, revenue generated) requires minimal processing but do not consider interrelations; (b) Regressions (e.g. correlation between R\&D spending and GDP value added) work by defining and correlating pairwise data, for example, on innovation input-outputs. Attribution is powerful in highly controlled settings but intermediary outputs are not considered and randomization is usually difficult in innovation settings. For statistical analysis to work, these pairwise data are usually disjoint and independent; (c) Networks (e.g. collaboration, spillover networks) may collect, map, and connect data of all relevant inputs, intermediaries, outputs, and outcomes, enabling non-Euclidean analysis that is scalable to innovation’s complexity, emergence, and pathways. Challenge is usually the lack of robust relational data and network analytical packages.}
    \label{fig:network_vs_others}
\end{figure}

R\&D programs \emph{configure} heterogeneous people, places, and things into new arrangements, which, in turn, set the conditions for \emph{communication} to propagate old and new ideas across the innovation system. Despite configurational and communication feedback loops being acknowledged as important \cite{RN276,RN285}, they are not captured by standard statistical treatment of innovation which consider innovators and innovations in isolation: bibliometrics and scientometrics, alone, consider innovation variables as unrelated entities (Fig.\ref{fig:network_vs_others}a); regression models correlate many pairs of unrelated variables (Fig.\ref{fig:network_vs_others}b). For these  innovation statistics to establish causality, it is usually desirable that the data points and data pairs be disjoint and independent \cite{RN3210}. Since the seminal work by \cite{RN997}, network science emerged as a complementary means to interrogate more complex relationships. For example, co-authorship networks are now standard tools for mapping collaboration and identifying central researchers or institutions \citep{Glanzel2004,Breschi2010,Sampaio2016,Fagan2018}, while co-patenting networks illuminate technology development \cite{Doblinger2019,Shen2025}. Parallel work in science and technology studies exploit networks to characterize national and international collaboration patterns and their link to research impact \citep{Ortega2014,IsfandyariMoghaddam2023,Pepe2024}. Nevertheless, traditional innovation networks are typically constrained by the use of univariate data (e.g. patent only) \cite{RN3211,RN3214}, an issue recently rectified by the use of multilayer networks to measure multivariate feedback loops \cite{RN2631,RN2871} .

{Existing innovation-network studies typically privilege one relation type at a time: co-authorship, co-invention, patent citation, grant funding, topic similarity, or institutional collaboration. These views are valuable but partial. R\&D programs are inherently typed systems \cite{RN274,RN182}: people link to organizations, organizations to projects, projects to outputs, outputs to funders, and outputs to other outputs through citation and patent relations. Collapsing these layers loses the architecture that program managers actually design. Our contribution is to preserve that heterogeneity while keeping the program as the organizing unit of analysis.}

{This study builds on that literature by portraying R\&D programs in three ways. First, we preserve a heterogeneous, typed directed graph that distinguishes people, places, and things and retains multiple relation types (funding, performance, affiliation, authorship, citation) rather than collapsing them into a single co-occurrence layer. Second, we take the formal R\&D program as our organizing unit, constructing bottom-up induced networks from project disclosures and composing them into an agency-wide view. Third, we use typed local patterns, drawing on work in systems biology and multiplex networks \citep{Milo2002,Bao2022,Tran2015,Takes2018}, to summarize recurring structures of configuration and communication in explicitly structural terms.}

\subsection{Network construction}

A \emph{network} (graph) is a mathematical object that represents entities as \emph{nodes} and their relationships as \emph{edges}. In our innovation network, \textbf{nodes} include \emph{people} (researchers, program directors), \emph{places} (research institutions, funding organizations, labs, firms, agencies), and \emph{things} (grants, papers, patents, and other technical outputs). \textbf{Edges} encode the relations between nodes; in our context these include funding, performance, affiliation, authorship, citation, and assignment.

{Table~\ref{tab:folk_knowledge} provides the inclusion criteria for our nodes and outlines qualitative hypotheses we can examine about challenge-led R\&D. Our approach uses a \emph{typed, directed graph with temporal attributes} so that we preserve direction (who funds whom, who cites whom), semantics (what kind of relationship), time (before/after the award via node-level dates), and the ability to compose program-level networks into an agency-wide view with shared nodes. We refer to the resulting overlap regions as ``communicating'' subgraphs, while treating them as structural overlap rather than direct observation of communication.}

{By \emph{topology}, we mean the patterned arrangement of node classes and relation types within and across programs: who and what is connected, through which relation, at what level of the program hierarchy, and with what recurring local structures. In this sense, R\&D programs are not generic collaboration networks. They combine administrative hierarchy (agency-program-project), capability assembly, output production, knowledge linkage, institutional anchoring, and cross-program overlap. Their topology is partly designed by managers and partly emergent through publications, patents, citations, affiliations, and co-funding relations.}

{Unlike many canonical network systems, R\&D program networks are explicitly managerial objects. Their topology is shaped by program calls, performer selection, staged funding, institutional participation, technical targeting, and output production. This makes topology not only an analytical property of the system, but also a potential object of design.}

While general formalisms for multilayer or multiplex networks exist \cite{RN1047,RN3140}, these have not been operationalized as empirical typed, directed graphs with temporal attributes for science-funding or innovation ecosystems. Our work therefore introduces a modelling approach that is rarely used in standard innovation studies or scientometric analyses. A key consequence of this formalism is that each R\&D program can be represented as an \emph{induced subgraph} of the global innovation network, preserving the full type and direction information of all edges whose endpoints belong to that program. This is a departure from conventional approaches that treat programs as independent project lists or aggregate them into bipartite funder-performer matrices. By composing these induced subgraphs, we can identify \emph{communicating subgraphs} (overlapping regions where programs share actors or outputs), revealing cross-program coordination channels that are invisible in project-level reporting.

\begin{table}[h]
\centering
\footnotesize
\caption{Nodes included in this study and their relevance to R\&D program structure}
\label{tab:folk_knowledge}

\setlength{\tabcolsep}{6pt}
\renewcommand{\arraystretch}{1.15}

\begin{threeparttable}
\begin{tabularx}{0.99\textwidth}{
    >{\raggedright\arraybackslash}p{0.14\textwidth}
    >{\raggedright\arraybackslash}p{0.26\textwidth}
    >{\raggedright\arraybackslash}X
}
\toprule
\textbf{Node type} &
\textbf{Sub-type / Inclusion criteria} &
\textbf{Observations about challenge-led R\&D that await measurement} \\
\midrule

``People'' &
\parbox[t]{\linewidth}{%
{\textbullet\ Researchers (aka.~performers)\\
\textbullet\ Principal investigators\\
\textbullet\ Program directors}%
} &
\parbox[t]{\linewidth}{%
\textbullet\ Program directors ``seed'', grow, and govern otherwise ``disparate'' social networks, pulling in new people to make radical innovations possible \cite{RN445}\\[2pt]
{\textbullet\ DARPA funds new networks whereas NSF funds the best people in existing networks \cite{RN448}}%
} \\
\addlinespace[0.4em]
\midrule

``Places'' &
\parbox[t]{\linewidth}{%
{\textbullet\ Universities\\
\textbullet\ Labs\\
\textbullet\ Firms\\
\textbullet\ Funding organizations}%
} &
\parbox[t]{\linewidth}{%
{\textbullet\ Coordinated investment of institutions are architected by program directors in multi-year R\&D programs designed to advance complex, large-scale ideas \cite{RN2771}\\[2pt]
\textbullet\ Other public-sector agencies and private funders may appear as recurrent co-funders or citation-linked funders around program outputs, providing structural evidence of overlap rather than direct observation of financial hand-offs \cite{RN445,RN2802}}%
} \\
\addlinespace[0.4em]
\midrule

``Things'' &
\parbox[t]{\linewidth}{%
\textbullet\ Papers\\
\textbullet\ Patents\\
\textbullet\ Grants\\
\textbullet\ Technical outputs\\%
\textbullet\ Programs\textsuperscript{1}\\
\textbullet\ Projects\textsuperscript{2}\\
} &
\parbox[t]{\linewidth}{%
\textbullet\ DARPA is often a first mover, derisking a niche technology through system demonstrators, whereas traditional funders focus on building established domains \cite{RN21}\\[2pt]
\textbullet\ Funding awards are characterized by portfolios and real options: funding rises and cuts mid-project are expected \cite{RN259,RN171}%
} \\
\bottomrule
\end{tabularx}

\begin{tablenotes}\footnotesize
    \item \textsuperscript{1}A program in this study refers to R\&D funding activities constructed by identifying specific technology needs or new capabilities; usually made up of a set of projects. \textsuperscript{2}A project is one of a portfolio of projects within a program with defined technical goals \cite{RN1282,RN21}.
\end{tablenotes}
\end{threeparttable}
\end{table}

\subsection{Formal definitions}
\label{sec:definitions}

\begin{definition}[Typed, directed graph with temporal attributes]
Let $V$ be a finite set of nodes, let $\mathcal{R}=\{\mathsf{funds},\mathsf{performs},\allowbreak\mathsf{performed\_by},\mathsf{affiliated\_with},\mathsf{authored},\mathsf{invented},\mathsf{cites},\mathsf{owns},\mathsf{produced},\mathsf{assigned}\}$ be a finite set of relation labels, and let $E\subseteq V\times V\times\mathcal{R}$ be a set of directed, labelled edges.
A \emph{typed, directed graph} is the tuple
\[
G=(V,E,\tau),
\]
where $\tau:V\to\{\mathsf{Person},\mathsf{Place},\mathsf{Thing}\}$ assigns each node a type. Each edge $e=(u,v,r)\in E$ connects source $u$ to target $v$ with relation $r$; because the relation label is part of the edge identity, the same ordered pair $(u,v)$ may participate in multiple edges with distinct labels, making $G$ a typed multi-graph. Temporal ordering is derived from node-level dates: each \textsf{Thing} node $z$ carries a timestamp $t(z)\in\mathbb{T}$ (publication year, patent filing date, or award interval), and edges inherit their temporal context from the dated nodes they connect. For nodes whose temporal extent is an interval (e.g.\ projects, grants), $t(z)$ denotes the interval $[t_{\mathrm{start}},t_{\mathrm{end}}]$; for nodes with a single date (e.g.\ publications), $t(z)$ is a point in $\mathbb{T}$. Affiliations are treated as static at the time of output: a person is affiliated with the institution recorded on the output at the date of that output. Co-authorship between two Person nodes is not stored as an explicit edge but is inferred when both share an outgoing $\mathsf{authored}$ or $\mathsf{invented}$ edge to a common output.
\end{definition}

\noindent Each coarse type $\tau(v)$ admits finer sub-types drawn from the ARPA-E impact documents and enriched via the Dimensions.ai bibliometric database~\cite{RN1696} and the Research Organization Registry (ROR):

\begin{itemize}
  \item \textbf{Person} sub-types: principal investigator (\textsf{pi}), co-author (\textsf{co\_author}), inventor (\textsf{inventor}), program director (\textsf{program\_director}), and co-awardee (\textsf{co\_awardee}). Author and inventor identities are resolved via Dimensions researcher IDs where available.
  \item \textbf{Place} sub-types: university (\textsf{education}), company (\textsf{company}), government agency (\textsf{government}), national laboratory or facility (\textsf{facility}), non-profit (\textsf{nonprofit}), and funding organization (\textsf{funder}). Organization types are assigned using the GRID/ROR type taxonomy; when a registry match is unavailable, names are classified heuristically.
  \item \textbf{Thing} sub-types: program (\textsf{program}), project (\textsf{project}), publication (\textsf{pub}), patent (\textsf{patent}), grant (\textsf{grant}), and cited publications/patents (\textsf{pub\_cited}, \textsf{patent\_cited}). Publication and patent metadata, including authors, research organizations, funders, supporting grants, forward/backward citations, and Fields of Research (FoR) categories, are retrieved from the Dimensions API.
\end{itemize}

\noindent We consider three complementary perspectives for constructing our people-places-things networks. 

\begin{definition}[Bottom-up induced networks]
For each project or program $p$, let $V_p\subseteq V$ be the set of people, places, and things associated with $p$ (seeded from ARPA-E impact documents and enriched as described in Section~\ref{ch:data}). The \emph{bottom--up network} of $p$ is the induced typed subgraph
\[
G_p = G[V_p],
\]
retaining all labelled, time-stamped edges whose endpoints both lie in $V_p$. These networks represent the local innovation structures orchestrated by individual programs.
\end{definition}

\begin{definition}[Communicating subgraphs]
Suppose $G_{p_1}=G[V_{p_1}]$ and $G_{p_2}=G[V_{p_2}]$ are two bottom--up program networks. Let $V_{p_1 p_2}:=V_{p_1}\cap V_{p_2}$ and define the overlap-induced subgraph
\[
C(p_1,p_2) := G[V_{p_1 p_2}].
\]
We call $C(p_1,p_2)$ a \emph{communicating subgraph} if $V_{p_1 p_2}$ contains at least one node of type $\mathsf{Person}$ or $\mathsf{Place}$. Communicating subgraphs represent overlaps in social or institutional actors through which distinct programs can exchange information, practices, or resources.
\end{definition}

\begin{definition}[Top-down composed network]
The \emph{top-down network} of the agency is the union
\[
G_{\mathrm{agency}} = \bigcup_{p\in\mathcal{P}} G_p,
\]
taken over all programs $\mathcal{P}$. Nodes and edges are merged if they represent the same underlying entity or relation. The resulting joint graph integrates program-level activities into a global innovation system view.
\end{definition}

{Intuitively, bottom-up induced networks expose how a single program is internally structured (Figs.~\ref{fig:amped_people}--\ref{fig:amped_things}), while top-down composition shows how programs relate to each other (Fig.~\ref{fig:arpae_all}). Communicating subgraphs are the bridges: they are the overlapping regions through which cross-program coordination or knowledge sharing could plausibly occur, even though the overlap itself does not prove that those processes took place.}

\begin{figure}[h!]
  \centering
  \includegraphics[width=0.95\textwidth]{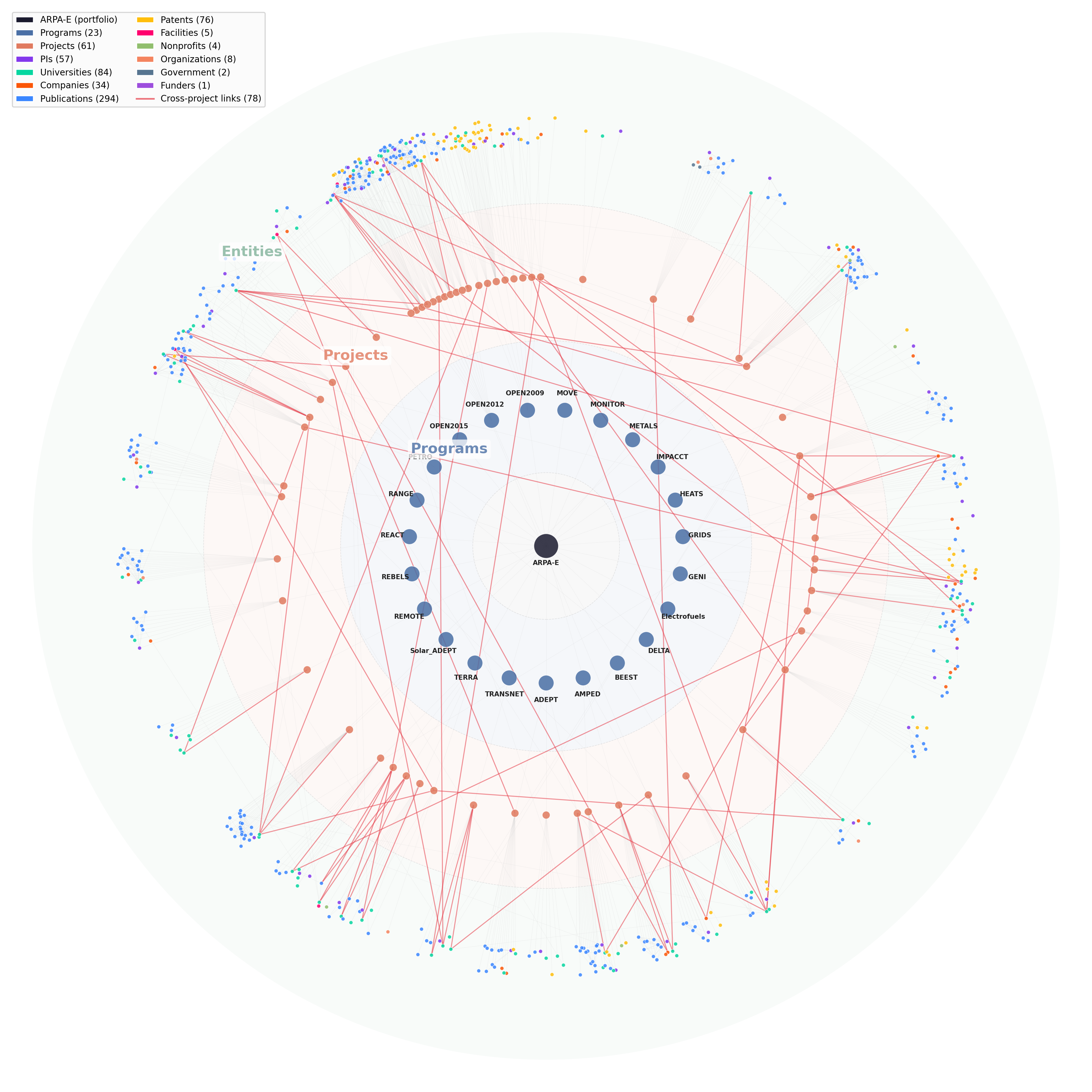}
  \caption{{\textbf{Ego network of the ARPA-E portfolio at radius~3.} The central dark node represents ARPA-E. Concentric shaded bands delineate the three hop layers: the inner band contains the 23 programs (blue); the middle band contains 61 projects (salmon); and the outer band contains 559 entities reached at three hops: 294 publications (blue), 84 universities (teal), 76 patents (gold), 57 PIs (purple), 34 companies (orange), and 14 other organizations and facilities. Red lines highlight 78 \emph{cross-project bridges}: edges connecting 32 entities (predominantly universities and national laboratories) that are shared across multiple projects. The figure visualizes structural overlap. This top-down view reveals the spread of influence from agency through programs and projects to the people, places, and things they mobilize, contrasting with the bottom-up induced program networks in Figs.~\ref{fig:amped_people}--\ref{fig:amped_things} that show internal program architecture.}}
  \label{fig:ego_portfolio}
\end{figure}

A complementary perspective is the \emph{ego network} \cite{RN1749}: the induced subgraph centred on a single actor at a given radius. Fig.~\ref{fig:ego_portfolio} shows the ego network of the ARPA-E portfolio node at radius~3, recovering the complete hierarchy from agency to programs to projects and onward to the people, places, and things each project mobilises. While the bottom-up induced program network (e.g., Fig.~\ref{fig:amped_people}) reveals the internal social and institutional architecture assembled by a single program, the top-down ego network reveals what a funder or portfolio manager ``sees'' from their vantage point: the full set of programs and projects under their oversight, and where they branch or converge. This distinction is analytically important: bottom-up views diagnose \emph{how} a program is configured; top-down ego views diagnose \emph{what} a funder's portfolio looks like as a connected structure. Program directors appear as nodes in the network when they are explicitly named in impact sheets (as the responsible manager of a program) or when they appear as co-authors or co-inventors on program outputs. However, they are not systematically linked to every project they oversee; future work with richer administrative data could construct ego networks centred on individual program directors to compare managerial styles at that finer scale.

\subsection{Network analysis}
\label{ch:motifs}

\subsubsection{Scoping network motifs}

{Once R\&D programs are represented as typed networks, local structural patterns can be interpreted as topological primitives of program architecture. We retain the term \emph{motif} for continuity with network-science usage, but the goal here is diagnostic rather than universal motif discovery. The patterns translate domain-relevant processes -- capability assembly, knowledge clustering, serial invention, co-funding, institutional linkage, company ownership, and output breadth -- into recurring graph structures.}

{We therefore identify two functional roles that R\&D program topology must capture: (i)~\textbf{configuration}, assembling heterogeneous actors and capabilities into productive arrangements internally~\cite{RN21,RN445}; and (ii)~\textbf{communication}, connecting program outputs, performers, and institutions to wider scientific, funding, and technological  \cite{RN1,RN182,RN960}. Configuration motifs capture how a program's internal actors are wired together, while communication motifs capture how a program's outputs and actors connect to external funders, institutions, and knowledge bases.}
\label{sec:knowledge_pickup}

\begin{figure}[h!]
  \centering

  \begin{subfigure}[t]{0.9\textwidth}
    \centering
    \includegraphics[width=\textwidth]{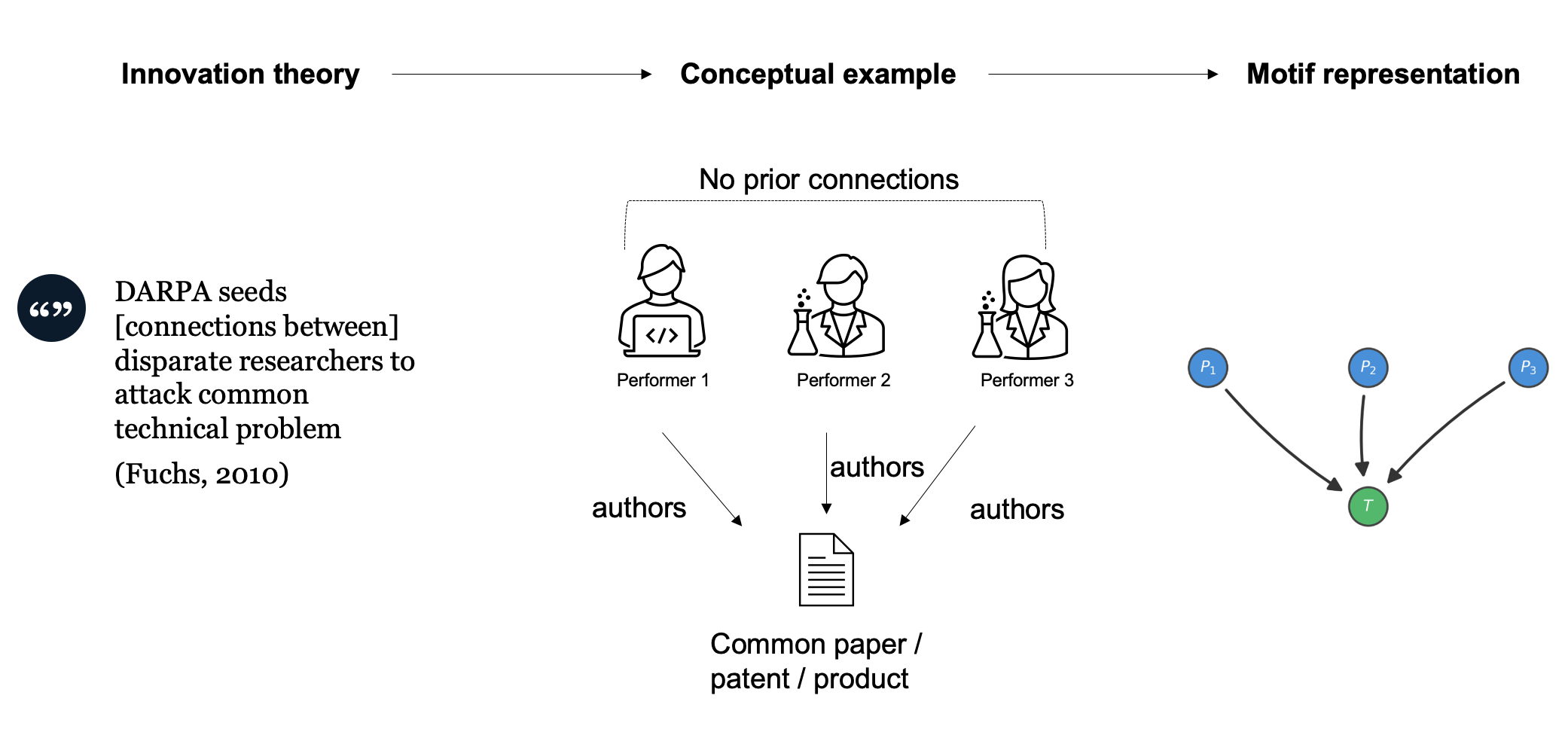}
    \caption{Collaboration motif.}
    \label{fig:motif_theory_a}
  \end{subfigure}
  \hfill
  \begin{subfigure}[t]{0.9\textwidth}
    \centering
    \includegraphics[width=\textwidth]{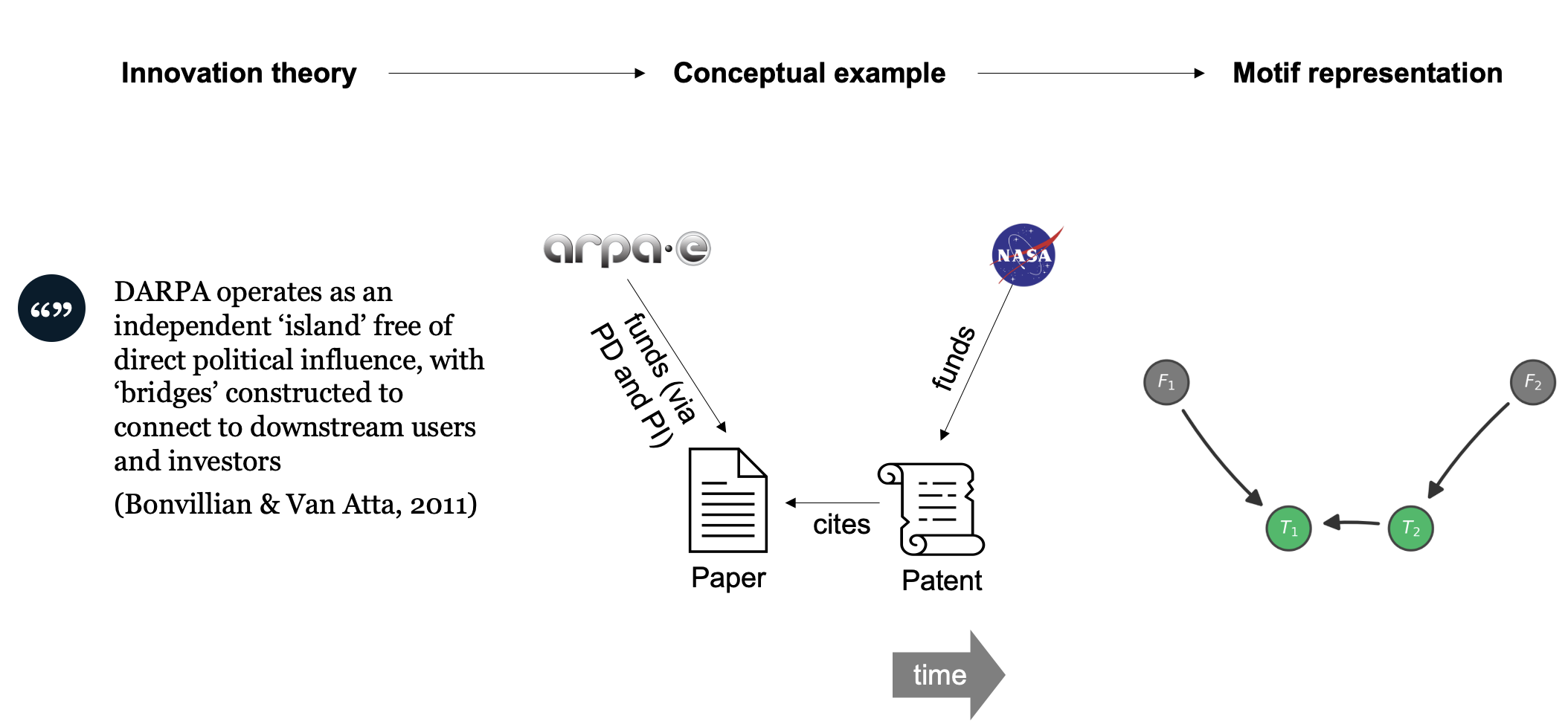}
    \caption{Follow-on funding motif.}
    \label{fig:motif_theory_b}
  \end{subfigure}

  \caption{\textbf{From innovation theory and mechanisms to network motifs.}}
  \label{fig:motif_theory}
\end{figure}

\begin{figure}[h!]
  \centering
  \includegraphics[width=0.95\textwidth]{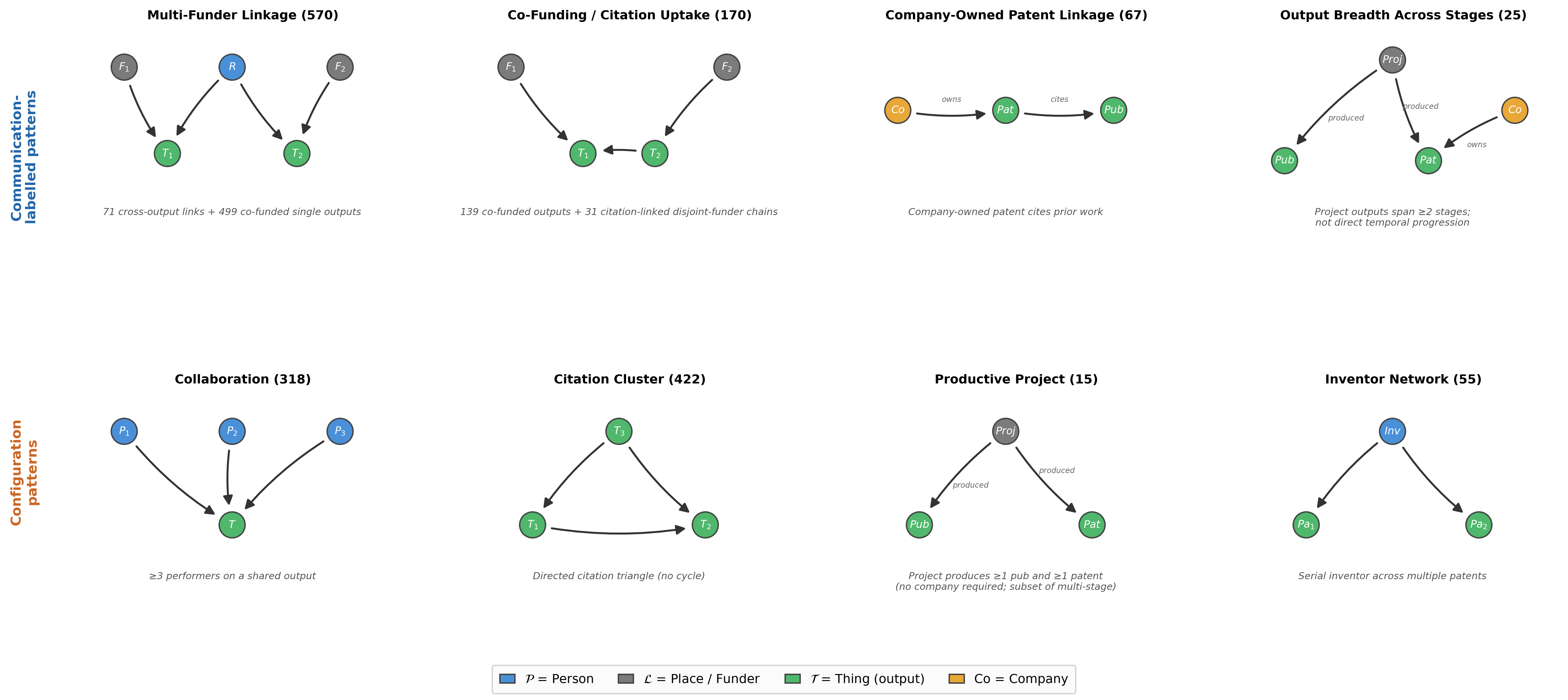}
  \caption{{\textbf{Network motifs considered in ARPA-E program induced networks are innovation theory-driven.}. Top row: four communication motifs that capture multi-funder, citation-linked, or company-linked structures around program outputs. Bottom row: four configuration motifs that capture how heterogeneous actors assemble around shared outputs. Each motif shows its type name and frequency across all 23 programs. All motifs are detected within program-induced subgraphs, so ARPA-E involvement is implicit in every instance. Edge labels on the three lifecycle-labelled motifs (\emph{company-patent linkage}, \emph{cross-stage output}, \emph{productive project}) show the semantic relation type (\textsf{owns}, \textsf{cites}, \textsf{produced}) that distinguishes them. The labels should be read as shorthand for the structural detector, not as direct observation of temporal stages or market outcomes.} For most motifs, the context chain is ARPA-E $\to$ Project $\to$ output(s), because all patterns are detected within the ARPA-E portfolio ego network (Fig.~\ref{fig:ego_portfolio} and Fig.~\ref{fig:motif_context}).}
  \label{fig:motif_inventory}
\end{figure}

{We pre-specify a bounded dictionary of eight typed local patterns (Fig.~\ref{fig:motif_inventory}). Several detectors in this inventory are theory-driven local configurations or degree-threshold patterns rather than canonical overrepresented subgraphs in the strict network-science sense. Their value is that they are domain-interpretable graph patterns: collaboration, citation clustering, productive projects, inventor networks, multi-funder linkages, co-funding, company-owned patent links, and cross-stage outputs. In typed, directed graphs the number of possible non-isomorphic subgraphs grows combinatorially with node and edge labels, so exhaustive enumeration is neither tractable nor especially informative here. Our approach is instead \emph{theory-driven}: we specify a small interpretable inventory of local structures that a reviewer or program analyst would recognize as relevant to program architecture, as illustrated in Fig.~\ref{fig:motif_theory}. These eight types are not meant to be permutationally exhaustive; they are chosen because each maps onto an interpretable local structure that ARPA-style agencies or evaluators may wish to monitor.}

\paragraph{Configuration motifs.}
{We identify four patterns that reflect how a program assembles its internal structure. A \emph{collaboration} motif consists of three or more co-authors or co-inventors on a single output, the most direct signal that a program assembles diverse human capital around one artefact. A \emph{citation cluster} is a closed triad in the undirected projection of the citation network (each pair of outputs connected by at least one directed citation edge), indicating a locally coherent knowledge cluster rather than isolated publications. Because directed citation cycles are removed in Step~2 of the detection pipeline, the triad may be a transitive directed triple (e.g.\ $A \to B$, $A \to C$, $B \to C$) rather than a cycle of reciprocal citations. A \emph{productive project} is a project node with outgoing \textsf{produced} edges to both at least one publication and at least one patent, capturing output breadth within a single funded effort. An \emph{inventor network} captures an inventor contributing to two or more patents, representing serial inventive activity within the program.}

\paragraph{Communication motifs.}
{We identify four patterns that reflect how knowledge, funding, and outputs relate to actors beyond a single project boundary. A \emph{{multi-funder linkage}} motif occurs when a researcher (PI, co-author, or inventor), traced through outgoing \textsf{authored} or \textsf{invented} edges, or an organization (university or company), traced through incoming \textsf{performed\_by} edges, has work funded by two or more distinct funders. The detector therefore captures shared or repeated funder association. \emph{{Co-funding linkage}} is detected in two ways: (a)~a single output (publication, patent, or project) that receives \textsf{funds} edges from two or more distinct funders, which is best read as co-funding; and (b)~a citation chain in which a program publication cites another publication with an entirely disjoint funder set, which we treat as a structural proxy for possible downstream uptake by a new funder rather than direct evidence of a new award. A \emph{{company-patent linkage}} motif is a three-node chain in which a company \textsf{owns} a patent that \textsf{cites} any prior work (publication, cited publication, or another patent). Finally, \emph{{cross-stage output}} is detected in two ways: (a)~a \emph{project} node whose \textsf{produced} outputs span at least two of the three stages: research (publication), development (patent), and commercial positioning (company \textsf{owns} the patent); and (b)~a three-node chain in which a company-owned patent \textsf{cites} an ARPA-E program publication. This detector should be read as output breadth across stages, not direct observation of a temporal progression through stages.}

{The three lifecycle-labelled patterns are thus ordered by restrictiveness (Fig.~\ref{fig:motif_inventory}): \emph{company-patent linkage} (67 instances) is the broadest, \emph{cross-stage output} requires a project whose outputs span $\geq 2$ innovation stages, and \emph{productive project} is the most restrictive. These should be read as overlapping detectors rather than exclusive lifecycle states.\footnote{Three counting notes: (i)~Path~(b) of cross-stage output found zero instances in the data; all cross-stage instances come from path~(a). (ii)~\emph{Productive project} (project $\xrightarrow{\textsf{produced}}$ pub \emph{and} project $\xrightarrow{\textsf{produced}}$ patent) is a strict subset of cross-stage path~(a); all productive-project instances are therefore also counted as cross-stage instances. (iii)~\emph{Company-patent linkage} is not a subset of cross-stage output: it requires company $\xrightarrow{\textsf{owns}}$ patent $\xrightarrow{\textsf{cites}}$ prior work but no project node, whereas cross-stage output requires project $\xrightarrow{\textsf{produced}}$ outputs spanning multiple stages. The eight motif types are not mutually exclusive and the reported total of 1,642 includes the productive-project/cross-stage overlap of 15 instances.}}

To formalize how knowledge propagates at the individual level, we define the notions of \emph{exposure} (when an actor first encounters program knowledge) and \emph{adoption} (when they incorporate it into new work). These definitions, together with formal definitions of induced subgraphs and ego-networks, are given in Appendix~\ref{app:definitions}. In this study, exposure and adoption are operationalized indirectly through motif counting; direct measurement of individual-level knowledge-pickup events is left to future work.

\subsubsection{Detecting network motifs}

We implement motif detection in Python using \texttt{NetworkX}. For each of the 23 programs, the algorithm proceeds in four stages.

\paragraph{Step 1: Graph construction.}
Load the program's combined people--places--things graph $G_p = (V_p, E_p, \tau_p)$, where nodes carry a type label (person, place, or thing sub-type) and each edge $(u,v,r)\in E_p$ carries a semantic relation label $r$ (\textsf{funds}, \textsf{performs}, \textsf{authored}, \textsf{invented}, \textsf{cites}, \textsf{owns}, \textsf{produced}, etc.).

\paragraph{Step 2: DAG enforcement.}
Remove citation cycles among \textsf{Thing} nodes by identifying strongly connected components in the citation subgraph and deleting intra-component \textsf{cites} edges, ensuring citation paths are acyclic.

\paragraph{Step 3: Pattern matching.}
Search for each of the eight predefined motif types by examining local node neighbourhoods. Specifically: \emph{collaboration}, outputs with $\geq 3$ incoming \textsf{authored} or \textsf{invented} edges; \emph{citation cluster}, all triangles in the undirected projection of the \textsf{cites} subgraph; \emph{productive project}, project nodes with outgoing \textsf{produced} edges to both a \textsf{pub} and a \textsf{patent}; \emph{inventor network}, inventor nodes with $\geq 2$ outgoing \textsf{invented} edges; \emph{multi-funder linkage}, for researcher nodes (\textsf{pi}, \textsf{co\_author}, \textsf{inventor}), follow outgoing \textsf{authored} or \textsf{invented} edges to outputs and collect their funders; for organization nodes (\textsf{education}, \textsf{company}), follow incoming \textsf{performed\_by} edges to the same end; a motif is recorded when $\geq 2$ distinct funders are found; \emph{co-funding linkage}, (a)~outputs with $\geq 2$ incoming \textsf{funds} edges from distinct funders, and (b)~citation chains where the citing and cited publications have different funders; \emph{company-patent linkage}, company-owned patents (\textsf{owns} edge from a \textsf{company} node) with at least one outgoing \textsf{cites} edge to any prior work; and \emph{cross-stage output}, (a)~project nodes whose \textsf{produced} outputs span $\geq 2$ of the three stages (publication, patent, company ownership), and (b)~a three-node chain where a company-owned patent \textsf{cites} an ARPA-E program publication (category \textsf{pub} only, not external \textsf{pub\_cited}).

\paragraph{Step 4: Aggregation.}
Count instances per motif type per program and concatenate into a cross-program summary.

\section{Data} \label{ch:data}

Data sources for network construction and motif detection include: (i)~ARPA-E project impact sheets (2010--2020, all 23 programs, 61 projects); (ii)~Dimensions.ai bibliometric data (publications, patents, citations); (iii)~Research Organization Registry (institution types and affiliations); (iv)~funding acknowledgement metadata. 

Existing assessments of ARPA-E have largely used these impact sheets to construct project-level outcome variables – counting publications, patents, and follow-on funding to estimate incremental impact relative to other US energy programs \citep{NAS2017_ARPAE,Pless2020,WhatWorksRDI2024}. We instead treat each sheet as a seed for reconstructing a local innovation network. Every declared output becomes a node that we enrich with bibliometric and funding metadata, linking it to co-authors, co-applicants, citing and cited documents, and acknowledging funders. This approach lets us trace how ARPA-E projects perturb their immediate neighborhood in the wider innovation system, rather than merely asking whether they produce “more” outputs. Our workflow draws on recent efforts to combine linked publication–grant–patent datasets for energy innovation studies \citep{Wilson2018,Dyevre2024}, but shifts the unit of analysis to the program – the level at which ARPA-style agencies design and exercise agency.

From each project in ARPA-E’s impact documents, we were able to extract basic metadata such as its technical scope, duration, and award size. Crucially for our network analysis, we can also extract the identity of program managers, project performers and their institutions, and project outputs in terms of journal publications and patents. We used these extracted information to identify the R\&D performers and institutions directly funded by ARPA-E, as well as the second and third degree connections to these entities to study how ARPA-E’s programs perturb its proximal innovation system. The analysis was conducted using a series of structured steps as outlined in Section \ref{ch:methods}.

For consistency, a \emph{program} is a set of R\&D funding activities constructed by identifying specific technology needs or new capabilities by program directors; a \emph{project} is one of a portfolio of projects within a program with defined technical goals, timeframe, and budget and executed by performers \cite{RN1282,RN21}.

\section{Results}
\label{ch:results_visuals}

\subsection{Structural overlaps and institutional bridges}

ARPA-E programs are designed as modular interventions targeting distinct technical risks, yet their system-level effects may depend on cross-program interactions that are not directly observable from administrative records. From public disclosures, ARPA-E projects are nested within programs, which are, in turn, nested under four themes: electricity generation, electricity grid and storage, efficiency and emissions, and transportation and storage. Given challenge-led R\&D structures programs solve different parts of a broader technical challenge while diversifying technical risks across performers \cite{RN259}, we examine whether projects and programs exhibit previously unobserved cross-talk. At the program level, some connections are identifiable directly from the project impact sheets \cite{RN2975,RN2973,RN2974}. For instance:

\begin{itemize}
    \item ADEPT, GENI and Solar ADEPT: connected through Tim Heidel as the common program director
    \item GRIDS and REACT: connected through Mark Johnson as the common program director
    \item Electrofuels and REMOTE: connected through Ramon Gonzalez as the common program director
\end{itemize}

At the project level, however, cross-program connectivity is largely invisible without network reconstruction. Fig.~\ref{fig:ego_portfolio} reveals {78 cross-project edges linking 32 bridge entities across 44 projects}, connected primarily via shared external institutions and, in one case, a shared co-funder. These connections are only recoverable via the multilayered network representation described in Section~\ref{ch:methods}. For instance:

\begin{itemize}
    \item ADEPT and Electrofuels: connected through the Massachusetts Institute of Technology, which hosts performers across multiple laboratories.
    \item IMPACCT and OPEN2012: connected through Eric Masanet and four shared institutions including the University of California, Berkeley and Lawrence Berkeley National Laboratory.
    \item PETRO and TERRA: connected through Steven Long, a principal investigator active in both programs (PETRO 2012--2017, TERRA 2015--2019).
\end{itemize}

{The recovered overlaps reveal a structurally sparse but nontrivial integration pattern. We report overlap at the program level throughout to keep denominators comparable. Only 15 of 1,056 researcher nodes (1.4\%) participate in more than one program, while 62 of 324 organization nodes (19.1\%) recur across programs when recurrent co-funders are included (34 of 245 non-funder organizations, or 13.9\%, when co-funders are excluded). Universities dominate this overlap set, while national laboratories are the most connective category proportionally, with 3 of 5 facilities linking multiple programs. The most frequent bridging institutions include MIT and Lawrence Berkeley National Laboratory (each spanning at least five programs) and the University of Illinois Urbana-Champaign (six programs). Publications rarely bridge programs directly (1 of 294 publication nodes in Fig.~\ref{fig:ego_portfolio}), so the observed overlap is concentrated in organizations rather than shared outputs.}

{At the program level, connectivity is widespread but uneven. 20 of the 23 programs (87.0\%) connect to at least one other program through shared entities, producing 44 unique program pairs linked by common performers or organizations. Some programs (e.g., TERRA, AMPED) exhibit dense external connectivity, whereas others (e.g., ADEPT, REBELS, MOVE) remain comparatively isolated. These patterns show recurrent overlap, but they should be interpreted cautiously: a shared university or laboratory node is a plausible channel for communication, not direct evidence that communication occurred. Fig.~\ref{fig:arpae_all} further shows recurrent overlap with other DOE organizations (15/23 programs), NSF's Directorate of Engineering (7/23 programs), and Lawrence Berkeley National Laboratory (7/23 programs). DARPA, MIT, the University of Illinois Urbana-Champaign, Brookhaven National Laboratory, the Office of Naval Research, and Oak Ridge National Laboratory also appear in at least five of the surveyed programs. These recurring organizations are best interpreted as structural anchors in the disclosed portfolio rather than direct measures of organizational memory or managerial orchestration.}

{Overall, the representation makes otherwise hidden overlap visible across nominally independent projects and programs. Those overlaps are structural proxies for possible cross-program interaction, not direct observations of tacit knowledge transfer. Even so, they matter descriptively: recurrent overlap can reveal shared institutional dependence, potential coordination opportunities, and limits to portfolio diversification that do not appear in standard project-level reporting.}

\subsection{Quantifying structural signatures and cross-program heterogeneity}
\label{ch:results_motif}

{Across the 23 ARPA-E program graphs, network size ranges from 33 to 682 nodes (median 113; Table~\ref{tab:summary_stats}). As reported in Section~\ref{ch:results_visuals}, cross-program overlap is concentrated in organization nodes (62 of 324, 19.1\%) rather than researcher nodes (15 of 1,056, 1.4\%), with all denominators computed at the program level from the combined portfolio graph.}

\paragraph{Structural composition of R\&D programs.}
{Using methods from Section~\ref{ch:motifs}, we construct a quantitative inventory of local structural patterns and detected 1,642 pattern instances across all programs: 810 configuration patterns (49.3\%) and 832 communication patterns (50.7\%).\footnote{All motif counts reported in this paper are computed programmatically from the per-program motif detection pipeline and cross-validated against the aggregate summary file (\texttt{Fig3e\_innovation\_motifs\_summary\_ALL.csv}). The eight motif type totals, multi-funder linkage (570), citation cluster (422), collaboration (318), co-funding linkage (170), company-patent linkage (67), inventor network (55), cross-stage output (25), productive project (15), sum to 1,642 and reproduce the communication/configuration split of 50.7\%/49.3\% exactly. Note that motif types are not mutually exclusive: all 15 productive-project instances are also counted among the 25 cross-stage instances (see footnote in Section~\ref{ch:motifs}). Co-funding linkage path~(b) counts only citation chains whose citing and cited publications have \emph{entirely disjoint} funder sets (31 instances); same-funder citation chains are excluded as they represent a program citing its own prior funding context rather than a new-funder linkage.} Although communication and configuration patterns are approximately evenly split (50.7\% vs.\ 49.3\%),  that split is primarily a bookkeeping summary of the chosen detector set. The null-model analysis (Section~\ref{sec:robustness}) indicates that most counts are largely explained by the network's typed degree distribution. {Citation clusters are the principal exception: 11 of 23 programs exceed a descriptive reference level of $z=2$ relative to the degree-preserving null. We use that benchmark only as a compact summary of unusually large positive null contrasts; it is not a p-value cutoff or a formal significance test.}

Given these detectors use different counting units (project, inventor, output, or citation triangle) and are not mutually exclusive, they serve as a measurement of overlapping structural features rather than a directly comparable set of event counts. Fig.~\ref{fig:motif_inventory} illustrates the pattern types alongside their frequencies.}

\begin{figure}[h!]
  \centering
  \includegraphics[width=0.97\textwidth]{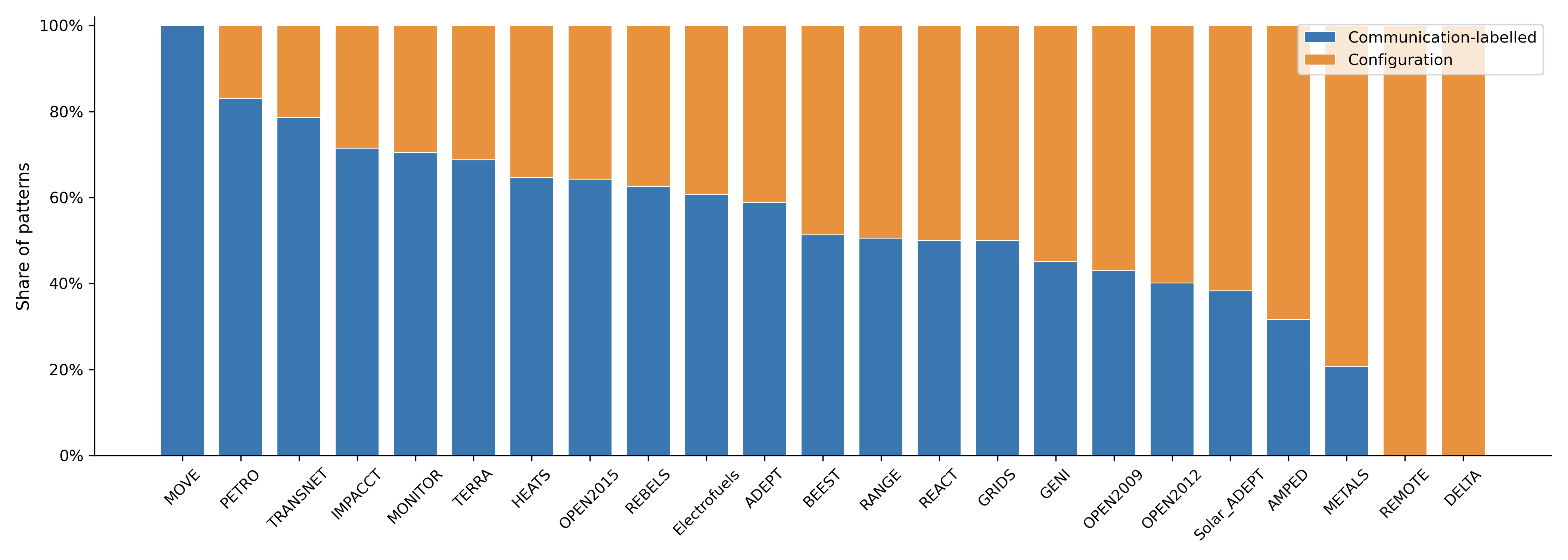}
  \caption{{\textbf{Pattern-category composition by ARPA-E program (2010--2020).} Stacked percentage shares showing the split between configuration patterns (collaboration, productive projects, inventor networks, citation clusters; orange) and communication patterns (multi-funder linkage, co-funding linkage, company-patent linkage, cross-stage output; blue). Programs are ordered by communication share (descending). Since the categories are theory-assigned and not mutually exclusive, the plot should be read as a descriptive summary rather than a discovered binary typology.}}
  \label{fig:motif_category_shares}
\end{figure}

\paragraph{{Detector overlap.}}
{Because multiple detectors can fire on the same underlying output nodes, the gross total of 1,642 instances overstates the number of independent structural events. Table~\ref{tab:detector_overlap} reports the pairwise instance-level overlap: cell $(A,B)$ gives the number of type-$A$ instances that share at least one output node (publication, patent, or project) with any type-$B$ instance. The largest overlap is between multi-funder linkage and co-funding linkage: 567 of 570 multi-funder linkage instances share an output with a co-funding linkage instance, because both detectors fire whenever an output has two or more funders. Citation clusters also share output nodes extensively with multi-funder linkage (366 of 422) and co-funding linkage (365 of 422), since highly cited outputs are disproportionately likely to be co-funded. We define a \emph{unique output node} as any publication, cited publication, patent, cited patent, or project node that participates in at least one motif instance; there are 576 such nodes across all programs. A \emph{unique output-node combination} is a (detector-type, output-node) pair, counting each output at most once per detector type. Overall, 322 of these 576 unique output nodes (55.9\%) participate in instances of two or more detector types. The 1,642 gross instances reduce to 1,247 unique output-node combinations when deduplicated, and the communication/configuration split is partly a consequence of the co-funding detectors counting the same multi-funder outputs through different structural paths. The heatmap, correlation matrix, and clustering analyses reported below use the gross per-type counts; readers should therefore interpret cross-type comparisons as descriptions of overlapping structural features rather than counts of independent events.}
{R\&D programs can nevertheless be compared at the level of measured structural composition by the share of each detected pattern. Fig.~\ref{fig:motif_category_shares} shows the relative share of configuration and communication patterns across ARPA-E programs. The overall split is near-even (50.7\% communication, 49.3\% configuration), but this balance should not be overinterpreted: the categories are theory-assigned, not empirically discovered, and one detector (\emph{productive project}) is a subset of another (\emph{cross-stage output}). The category plot is therefore best read as a structural summary of program architecture rather than a definitive classification. Collaboration accounts for 19\% of all detected patterns and citation clusters for 26\%, making these the two largest components of the inventory.}

\begin{figure}[h!]
  \centering
  \includegraphics[width=\textwidth]{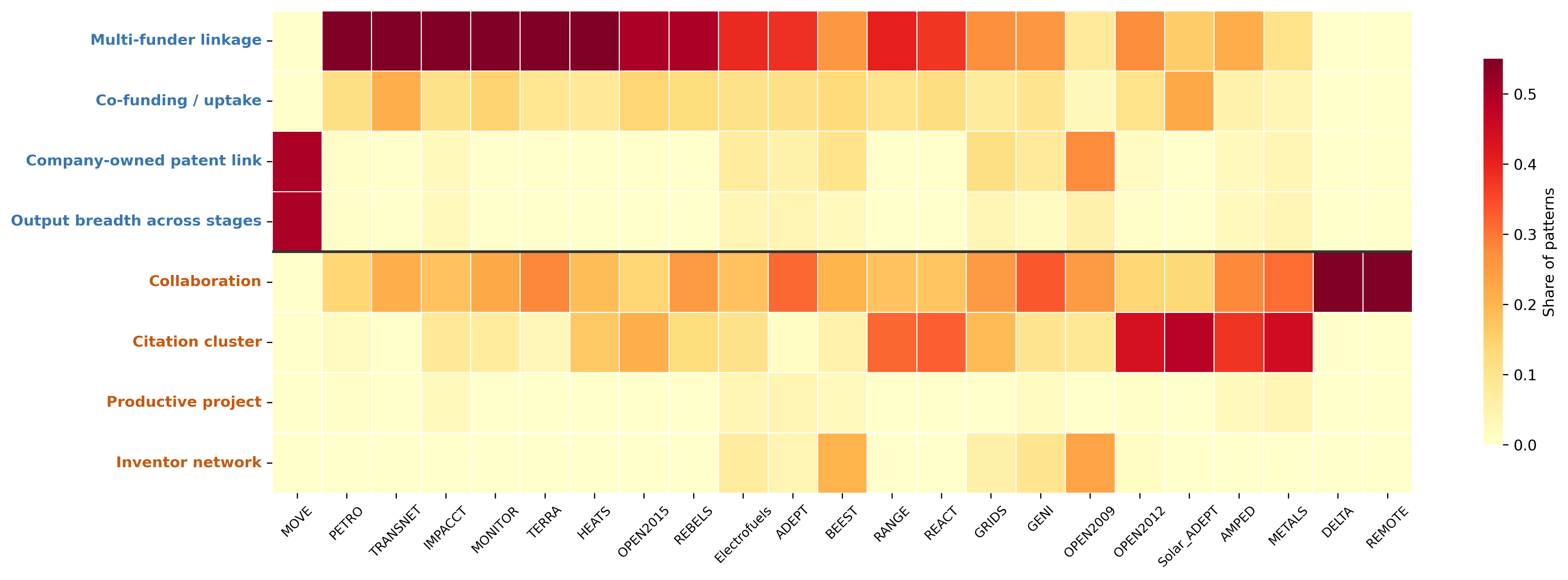}
  \caption{{\textbf{Pattern-type prevalence matrix across all ARPA-E programs (2010--2020).} Rows represent programs (ordered by communication share, top to bottom); columns show individual pattern types. Color intensity indicates the normalized prevalence (share of total detected patterns in that program). The labels \emph{multi-funder linkage}, \emph{co-funding linkage}, and \emph{company-patent linkage} reflect the structural detectors defined in Section~\ref{ch:motifs}.}}
  \label{fig:motif_type_heatmap}
\end{figure}

\begin{figure}[h!]
  \centering
  \includegraphics[width=0.69\textwidth]{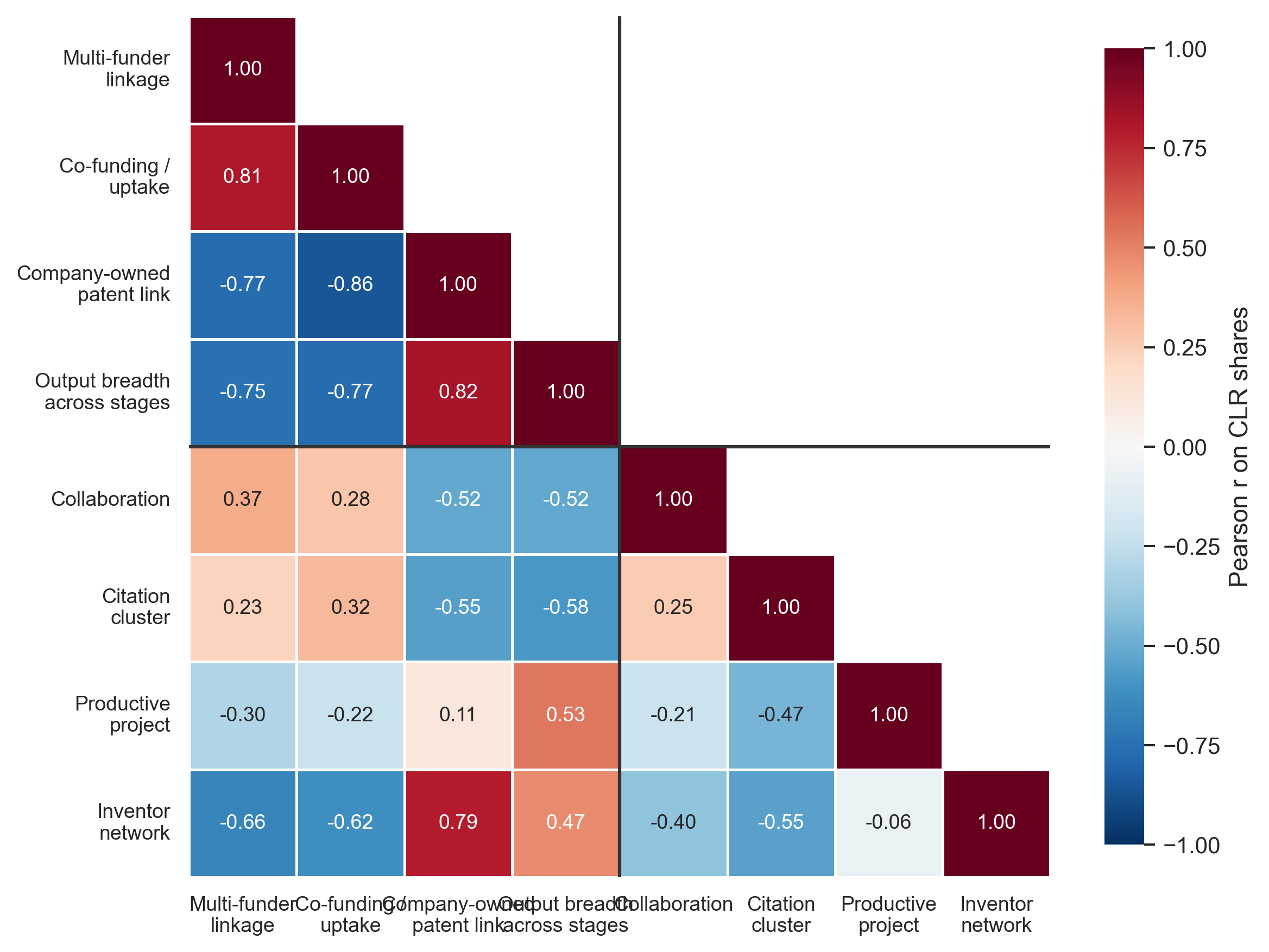}
  \caption{{\textbf{Association matrix between pattern types.} The heatmap summarizes pairwise associations between program-level pattern compositions.}}
  \label{fig:motif_correlation}
\end{figure}

\begin{figure}[h!]
  \centering
  \includegraphics[width=\textwidth]{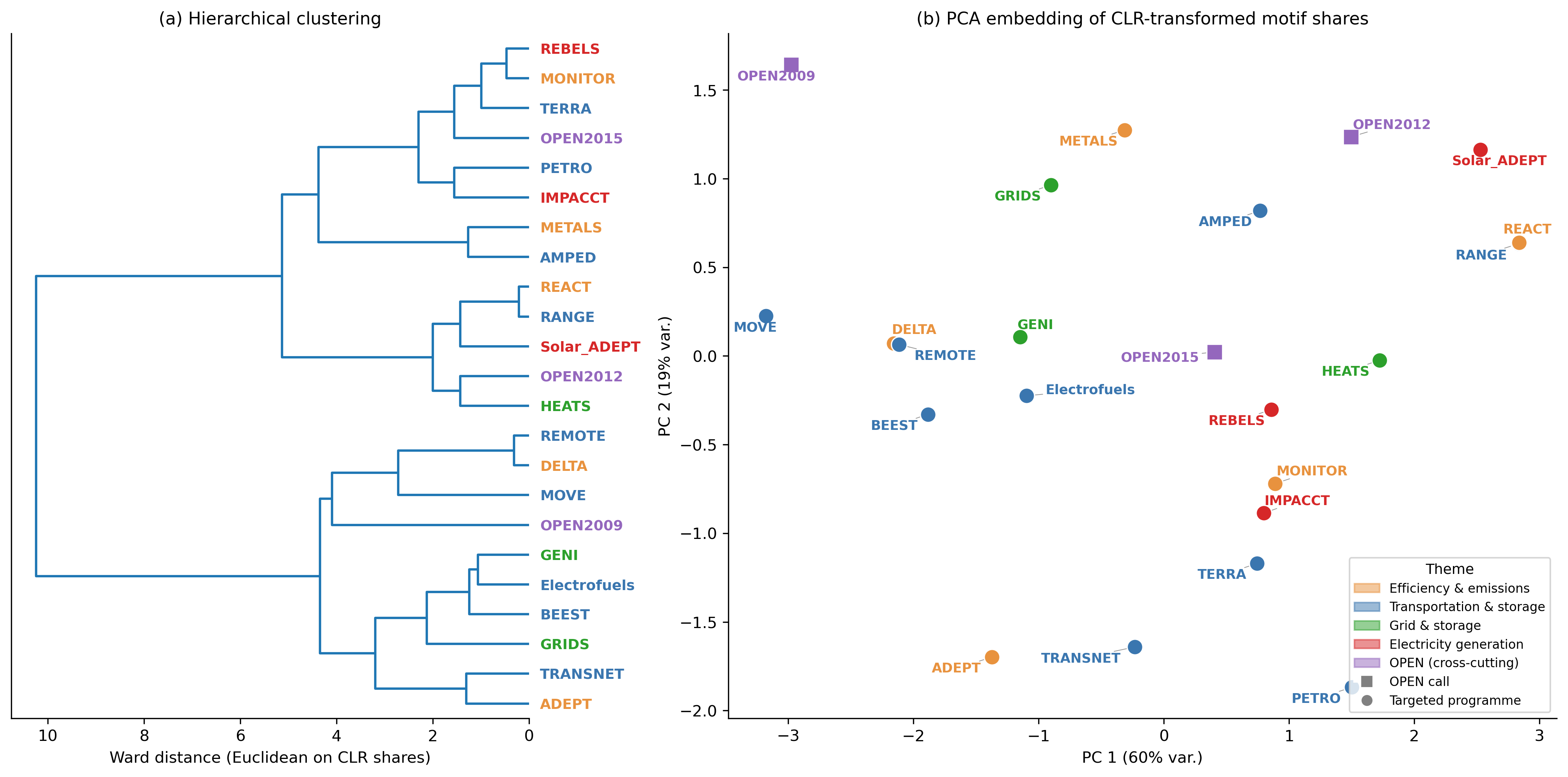}
  \caption{{\textbf{Program clustering by pattern fingerprints.} (a)~Hierarchical clustering of the 23 ARPA-E programs using Euclidean distance with Ward linkage on centered log-ratio transformed motif-share vectors. (b)~PCA projection of the same transformed vectors into two dimensions. Point colors and dendrogram label colors indicate ARPA-E thematic categories: orange = efficiency and emissions, blue = transportation and storage, green = grid and storage, red = electricity generation, purple = OPEN (cross-cutting). Squares denote OPEN calls; circles denote targeted programs. The figure is interpreted descriptively as broad convergence between structural fingerprints and thematic grouping, not as formal validation.}}
  \label{fig:clustering}
\end{figure}

\paragraph{Fingerprinting intra-agency R\&D programs.}
{Fig.~\ref{fig:motif_type_heatmap} shows the normalized distribution of all eight motif types per program,  providing a program-level signature of architectural composition. The detector labelled \emph{co-funding linkage} (170 total, across 20 of 23 programs) mixes 139 co-funded outputs with 31 citation-linked disjoint-funder chains; OPEN2012 (52 instances), Solar ADEPT (18), and PETRO (17) are the main concentrations. The detector labelled \emph{multi-funder linkage} is the single most common pattern overall (35\% of all patterns; 570 instances), but its composition is important: 499 of those 570 instances are single co-funded outputs rather than cross-output performer transitions.} 
{Fig.~\ref{fig:motif_correlation} summarizes pairwise associations between program-level pattern compositions. We treat these as structural covariation patterns rather than formal causal or inferential results, both because the sample is small ($n=23$ programs) and because motif shares are compositional. 

ARPA-E programs differ markedly in their pattern ``fingerprints'' even with near-equal aggregate totals (Figs.~\ref{fig:motif_category_shares} and \ref{fig:motif_type_heatmap}). {Among programs with at least five total motif instances,} PETRO carries a high share of communication-labelled detectors, while METALS carries a high share of collaboration and citation-cluster patterns. {Three programs (DELTA, MOVE, REMOTE) have fewer than five total motif instances and their share-based fingerprints are unreliable; we retain them in the figures for completeness but exclude them from prose-level interpretation.} Because program size varies substantially (from 33 to 682 nodes), normalized pattern proportions are more informative than raw counts for descriptive comparison. Hierarchical clustering of program pattern compositions (Fig.~\ref{fig:clustering}) also shows broad convergence with ARPA-E thematic groupings, which we treat as descriptive face validity rather than external validation in a strict statistical sense. We perform a robustness test in Appendix \ref{sec:robustness}.}

\section{Discussion}
\label{ch:discussion}
{This paper introduces a topology for R\&D program systems. Network science has long characterized the structure of transport, communication, biological, supply-chain, digital, and social systems. Mission-oriented R\&D programs are likewise relational, but their topology has remained difficult to observe because relevant entities are scattered across administrative records, publication databases, patent systems, funding acknowledgements, and institutional registries. By representing programs as typed networks of people, places, and things, we make this topology visible.}

{The contribution is therefore theoretical as well as methodological. We define the R\&D program as a network object, identify the program-induced subgraph as a meso-scale unit of analysis, and propose a dictionary of local structural patterns for describing how R\&D systems assemble capabilities and connect to wider knowledge and institutional environments.}

{This article makes three contributions. First, it contributes to network science by defining R\&D programs as typed, directed, program-induced networks rather than as lists of projects or single-layer collaboration graphs. This representation preserves the heterogeneity of people, organizations, funders, publications, patents, and projects while keeping the program boundary analytically visible. Second, it contributes to R\&D management by providing a practical way to inspect program architecture: who was assembled, which institutions anchored the work, what outputs were produced, and where those outputs connected beyond the program. Third, it contributes to innovation policy by showing how agencies can build a structural evidence base for mission-oriented R\&D, complementing output counts with maps of capability assembly, institutional overlap, and knowledge linkage.}

{For network science, the program-induced subgraph provides a meso-scale unit of analysis. It sits between micro-level collaboration networks and macro-level innovation systems, preserving the program boundary while retaining links to wider scientific, institutional, and technological neighborhoods. Figure~\ref{fig:ego_portfolio} is therefore not merely illustrative; it demonstrates the object created by the framework, moving from agency to programs, projects, publications, patents, universities, firms, PIs, and other entities.}

{For R\&D management practice, the framework provides a diagnostic map of program assembly. It allows managers to see which people, institutions, and outputs are connected inside a program; whether projects remain isolated or share organizational anchors; whether firms, universities, and laboratories occupy different structural roles; and whether outputs connect outward through citations, patents, and funders. These maps can support program design, mid-course review, and post-program learning without requiring the program to be reduced to a single performance score.}

{For program managers, the representation makes several architectural questions inspectable. These include whether a program concentrates activity around a few institutional anchors or diversifies across many independent performers; whether projects are connected through shared organizations, shared outputs, or shared technical neighborhoods; and whether firms, universities, and funders occupy distinct positions around patents, publications, and follow-on work. The AMPED program architecture reported in the appendix illustrates this managerial view at a recognizable scale: people, institutions, outputs, and citation neighborhoods can be inspected as one program structure rather than as separate lists.}

{For policy, the framework offers a way to build organizational memory across programs. Mission agencies often know individual programs well while they are running, but cross-program learning is harder: records are fragmented, staff rotate, and outputs mature over long time horizons. A typed network record can preserve how each program was assembled and how it touched other programs, institutions, and knowledge domains. This makes it possible to compare programs without forcing them into a single impact score.}

{Policy evaluation of mission-oriented R\&D should therefore include program architecture as an object of evidence. Agencies should be able to describe how programs assembled capabilities, where programs overlapped, which institutions recurred across missions, and how outputs connected to wider scientific and technological systems. Typed network records provide a practical way to produce that evidence.}

{For policy, the implication is that mission-oriented agencies should track not only outputs but also topology: the structure of capability assembly, institutional overlap, knowledge linkage, and cross-program connection generated by public R\&D programs.}

{Three empirical observations illustrate what the representation makes visible. First, cross-program overlap is mediated by a small number of recurring organizational anchors rather than widespread performer overlap: only 1.4\% of researchers appear in more than one program, while 62 of 324 organization nodes do, with MIT, Lawrence Berkeley National Laboratory, and the University of Illinois Urbana-Champaign among the most frequently recurring anchors. A concise policy reading is that ARPA-E programs diversify through people but reconnect through places. Second, program pattern fingerprints are heterogeneous and show broad convergence with ARPA-E thematic categories, suggesting that the typed pattern inventory captures meaningful variation in program architecture. Third, citation clusters are the only detector whose counts are repeatedly elevated relative to the degree-preserving null, with 11 of 23 programs above the descriptive reference level of $z=2$ ($n=100$ randomizations). This null-model contrast is useful supporting evidence, but it is not the center of the paper's contribution; it illustrates how the representation can support network-scientific comparison once the program object has been reconstructed.}

{Practically, the framework can support program design and program review. During program design, motif diagnostics can reveal whether portfolios assemble complementary capabilities or remain siloed across people, places, and technological outputs. During program execution, evolving network patterns may flag changing overlap, citation clustering, or new multi-funder linkages, enabling mid-course adjustment. After program completion, structural signatures provide a basis for comparing programs and documenting ecosystem structural impact beyond simple output counts.}

{A natural question for R\&D funders  is: what data would need to be collected routinely to appraise the architecture and ecosystem impact of R\&D programs? The minimum viable dataset corresponds closely to what mission-oriented agencies already record internally but rarely structure for network analysis \cite{RN171,RN3049}. At award, funders should log each performer (person and organization) with a persistent identifier (e.g.\ ORCID, ROR), the funding amount and source, and the declared outputs (publication DOIs, patent numbers, prototype descriptions). During execution, mid-course reports should capture any new co-funders, sub-contractor relationships, and inter-project personnel moves. At closure, a structured impact sheet recording all outputs with their bibliometric identifiers, citing documents, and any follow-on awards from non-ARPA-E sources directly populates the \textsf{funds}, \textsf{authored}/\textsf{invented}, \textsf{produced}, and \textsf{cites} edge types used here. Two additional fields that are currently missing from public records but would substantially enrich network-based, ecosystem impact analysis proposed in this study are: (i)~the identities and outcomes of \emph{unfunded} proposals submitted to the same program call, which would enable a counterfactual comparison group; and (ii)~the post-award career and funding trajectories of performing teams beyond the program window, which would allow direct measurement of performer mobility and follow-on funding over longer horizons. Agencies that begin collecting these fields systematically -- even in simple spreadsheet form -- would be able to reconstruct the full eight-motif profile reported here for any program in under a day of computation, and could track structural changes in real time as a program executes. The people-places-things schema is intentionally generic: it transfers directly to ARPA-H, ARIA, SPRIND, NIH study sections, or any agency that links performers, funders, and knowledge outputs, without modification to the detection code.}

Although this study analyzes the full corpus of publicly available structured project output data for ARPA-E at the time of writing, it is necessarily bounded by what is disclosed in project impact sheets, funding, and bibliometric databases (Table \ref{tab:summary_stats}), which are subject to bias to disclose successful and mature cases only and intrinsic data coverage issues. The sources are among the most comprehensive standardized records of contemporary mission-oriented R\&D agency at the time of writing, but they may omit informal knowledge transfer, proprietary outputs, tacit learning, and long-run market outcomes that unfold beyond the observable record. More broadly, the observable network captures only a subset of the actors, organizations, and knowledge artifacts involved in program ecosystems. Public records primarily document funded performers and formal outputs, but may omit downstream entrepreneurs, investors, adopters, supply-chain partners, trainees, infrastructure assets, and intermediate technological artifacts such as prototypes, software, standards, or data resources. The reconstructed structures should therefore be interpreted as conservative representations of ecosystem activity. Observed variation across projects and programs reflects differences in program scale, technology domain, maturity, collaboration norms, and time since award rather than measurement noise alone. Network representations also encode assumptions about relevance, attribution, and temporal ordering that require careful justification, and are most informative when relational structure -- rather than isolated inputs or outputs -- is central to the research question. Finally, this study also does not determine whether the network structures observed in ARPA-E programs are unique relative to other funding organizations, nor whether ecosystem effects persist over long time horizons.

Future research could apply the framework comparatively across agencies (for example, DARPA or NIH) and longitudinally as richer administrative data become available. In particular, improved program reporting and external case-based validations could enable triangulation by collecting information on pre- and post-award funding trajectories, collaboration histories, complementarities across projects, unfunded proposals, informal partnerships, and the career trajectories of participating teams, dimensions that are currently only partially observable in public records.

{Despite these limitations, the empirical foundation of this study is meaningful. We reconstruct networks from the complete set of disclosed ARPA-E project outputs over its first decade, using what is likely among the most comprehensive structured public datasets currently accessible for a contemporary mission-oriented R\&D agency. While incomplete by design, these data mirror the types of programmatic records routinely maintained within funding organizations. The people-places-things representation is also flexible: additional node types, relationships, or temporal windows can be incorporated as richer internal or proprietary data become available, making the framework transferable across institutional settings.}

{The growing prevalence of mission-oriented agencies, from ARPA-E and ARPA-H to ARIA, SPRIND, and philanthropic mission platforms \cite{RN3118,RN3024}, calls for evidence that captures how mission-driven R\&D programs are structured, not just what they produce. The framework developed here responds to long-standing calls for more system-level evaluation of R\&D programs \cite{RN2792}, offering a scalable complement to qualitative insight and traditional scientometrics.}

Back in 2005, then US National Science and Technology Council Director John Marburger acknowledged that “the complexity of the scientific endeavor is surpassing the ability of experts within particular scientific disciplines to understand its totality” \cite{RN3040}. He called for R\&D program evaluations to provide “scientifically rigorous, quantitative basis…[to] assess impacts of the…scientific and engineering enterprise, improve their understanding of its dynamics” \cite{RN2111}.

\section{Conclusion}
\label{ch:conclusion}

{Mission-oriented R\&D programs are designed relationally but often evaluated atomistically. By representing programs as typed networks of people, places, and things, this paper makes program architecture visible as an object of network analysis and managerial judgment. Applied to ARPA-E, the framework reveals programs as structured assemblies with distinct fingerprints and cross-program institutional bridges.}

{The broader contribution is a transferable representation: any agency that records performers, organizations, funders, and outputs can begin to map how its programs are assembled and connected. As mission-oriented R\&D agencies proliferate, this kind of structural evidence will become important for program design, portfolio learning, and policy evaluation.}

\clearpage

\appendix
\section{Formal definitions}
\label{app:definitions}

This appendix collects formal definitions referenced in the main text.

\begin{definition}[Exposure to program knowledge]
For a program $p$, let $\mathcal{K}_p \subset V$ be its \emph{knowledge set}: the \textsf{Thing} nodes (papers, patents, prototypes) produced during the award.
A person $u$ is \emph{exposed} to program $p$ at time
\[
T^{\mathrm{exp}}_{u,p} := \min \{\, t(k)\ :\ k\in\mathcal{K}_p \text{ and any of \emph{(E1)}--\emph{(E3)} holds} \,\}.
\]
\vspace{-0.8em}
\begin{itemize}
  \item[\emph{(E1)}] $u \xrightarrow{\mathsf{authored}/\mathsf{invented}} k$ (direct performer on a $p$ output);
  \item[\emph{(E2)}] $u$ and some $u'$ both have outgoing $\mathsf{authored}$ or $\mathsf{invented}$ edges to a common output, and $u' \xrightarrow{\mathsf{authored}/\mathsf{invented}} k$ for some $k\in\mathcal{K}_p$ (indirect via shared authorship/inventorship);
  \item[\emph{(E3)}] $u \xrightarrow{\mathsf{authored}/\mathsf{invented}} z$ and $z \xrightarrow{\mathsf{cites}} k$ for some $k\in\mathcal{K}_p$ (citation-based contact).
\end{itemize}
If multiple conditions hold, use the earliest time. Pre/during/post windows are defined relative to the award start~$T^{\mathrm{start}}_p$ and end~$T^{\mathrm{end}}_p$.
\end{definition}

\begin{definition}[Adoption (knowledge pickup)]
After exposure, $u$ \emph{adopts} knowledge from $p$ at the earliest time
\[
T^{\mathrm{adopt}}_{u,p} := \min \{\, t(z)\ :\ z \text{ is a \textsf{Thing} produced by } u,\; t(z) > T^{\mathrm{exp}}_{u,p},\; \text{and any of \emph{(A1)}--\emph{(A3)} holds} \,\},
\]
where:
\begin{itemize}
  \item[\emph{(A1)}] \textbf{Direct reuse:} $z \xrightarrow{\mathsf{cites}} k$ for some $k\in\mathcal{K}_p$;
  \item[\emph{(A2)}] \textbf{Collaborative reuse:} $z$ is coauthored with $u'$ where $T^{\mathrm{exp}}_{u',p} < t(z)$ and $u'$ is exposed to $\mathcal{K}_p$ via \emph{(E1)}--\emph{(E3)};
  \item[\emph{(A3)}] \textbf{Funded hand-off:} $z$ acknowledges support from a funder $f\neq \textsf{ARPA-E}$ and $z \xrightarrow{\mathsf{cites}} k$ for some $k\in\mathcal{K}_p$.
\end{itemize}
\end{definition}

\begin{definition}[Induced subgraph]
Given $G=(V,E,\tau)$ and $U\subseteq V$, the \emph{induced subgraph} on $U$ is
\[
G[U] \;=\; \big(U,\; E|_U,\; \tau|_U\big),
\]
where $E|_U = \{(u,v,r)\in E : u\in U,\, v\in U\}$ retains all labelled edges whose endpoints both lie in $U$, together with the temporal attributes of the connected nodes.
\end{definition}

\begin{definition}[Ego-network]
For a node $v\in V$ and radius $r\in\mathbb{N}$, the \emph{ego-network} of $v$ at radius $r$ is the induced subgraph on all nodes reachable within $r$ hops from $v$. We use the ego-network concept to scope the enrichment of program networks: starting from each program's seed nodes (performers, outputs), we expand to $r\in\{1,2,3\}$ hops to capture co-authors, citing/cited works, and their funders and institutions. In practice, this expansion is performed during graph construction (Section~\ref{ch:data}) by querying Dimensions.ai for the immediate neighbourhood of each seed output.
\end{definition}
\clearpage

\section{Verification of network motifs}

\begin{figure}[h!]
  \centering
  \includegraphics[width=0.95\textwidth]{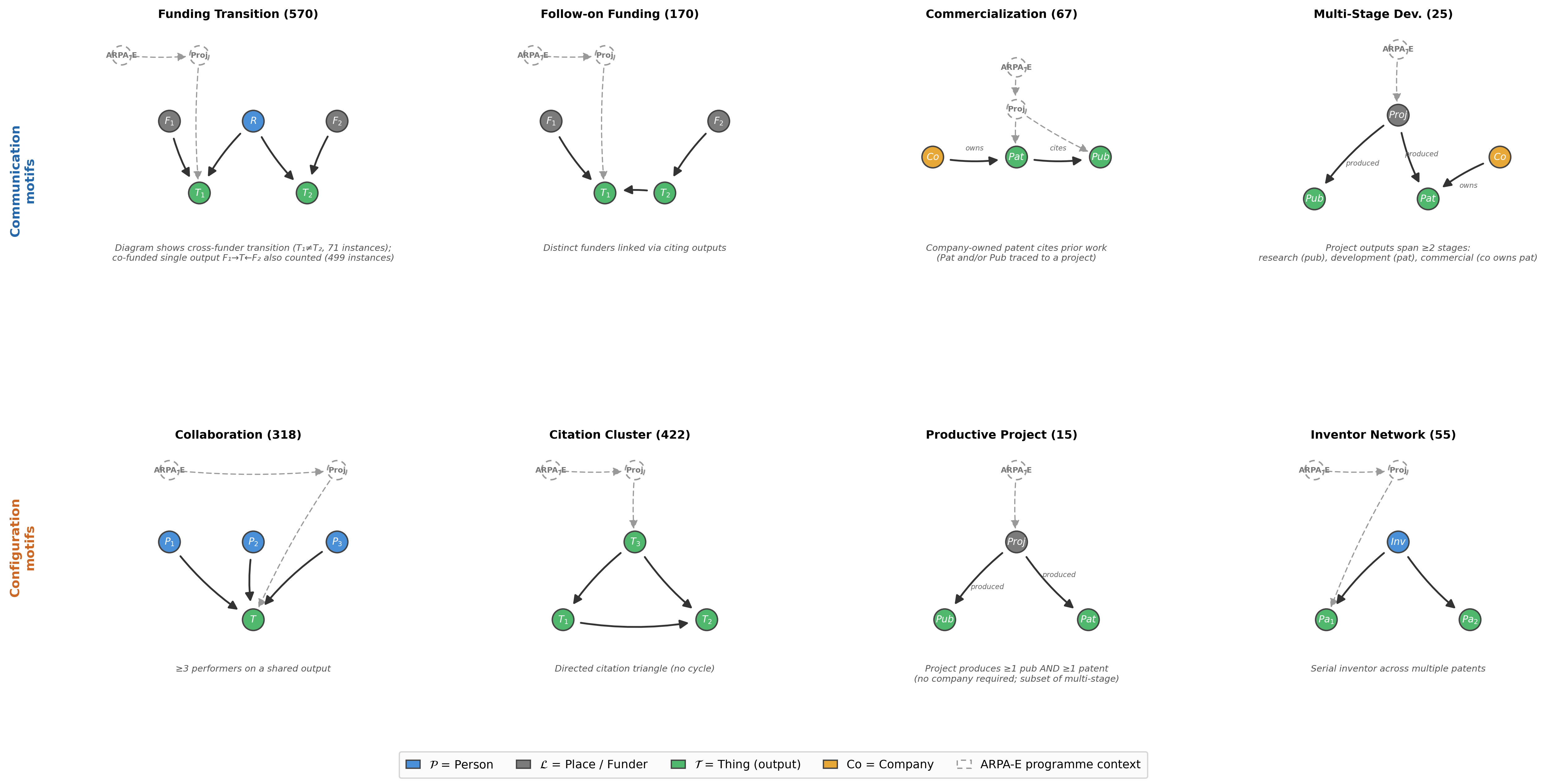}
  \caption{\textbf{Program context for each motif.} The same eight motifs as Fig.~\ref{fig:motif_inventory}, now annotated with dashed nodes and edges showing how each pattern is embedded in the ARPA-E program structure. For most motifs, the context chain is ARPA-E $\to$ Project $\to$ output(s). For motifs that already contain a project node in the solid pattern (cross-stage output, productive project), only the upstream ARPA-E $\to$ Project link is added. For company-patent linkage, the detection code does not explicitly check for a project node, but the patent and/or cited work must be traceable to a project in order to appear in the program-induced subgraph; the dashed arrows to both Pat and Pub indicate that either or both may be project outputs.}
  \label{fig:motif_context}
\end{figure}

A natural question is why the ARPA-E and project nodes are not included in the motif definitions themselves, given that every instance is connected to a project by construction. We exclude them deliberately for three reasons. First, the program-induced subgraph (ARPA-E $\to$ program $\to$ project $\to$ outputs) is \emph{identical scaffolding} for all eight motifs; including it would append the same chain to every pattern without adding discriminating information. Second, keeping motifs minimal, defined only by the structurally distinctive relationships (co-authorship, citation, ownership, funding), which preserves comparability: a three-node collaboration motif and a three-node company-patent linkage motif can be counted and compared directly, whereas appending project context would inflate all motifs by two to three nodes with no analytical gain. Third, the degree-preserving null model (Section~\ref{sec:robustness}) randomizes edges within the program subgraph; motifs that embed the fixed project hierarchy would be trivially preserved under randomization, confounding the enrichment test. The exogenous context is instead shown separately in Fig.~\ref{fig:motif_context}, making the program embedding transparent without conflating it with the motif structure.

Although not every motif definition explicitly checks for a project node, a post-hoc verification confirms that all 1,642 motif instances contain at least one node connected to a project via a \textsf{produced} edge; no orphan instances exist (Fig.~\ref{fig:motif_context}). In each motif the project-connected elements are the \emph{output nodes} (publications, patents):
\begin{itemize}[nosep]
  \item \emph{Multi-funder linkage} (570): the output node(s) are project publications in every instance. The motif covers two structural sub-patterns: (a)~a single output co-funded by two distinct funders ($F_1 \to T \leftarrow F_2$, $R \to T$), which accounts for 499 of 570 instances (87.5\%); and (b)~a performer who has authored different outputs under different funders ($F_1 \to T_1 \leftarrow R \to T_2 \leftarrow F_2$, $T_1 \neq T_2$), which accounts for 71 instances (12.5\%). The inventory figure (Fig.~\ref{fig:motif_inventory}) depicts sub-pattern~(b); both are detected by the code.
  \item \emph{Co-funding linkage} (170): in path~(b) (31 instances), both the citing and cited publications are project outputs; in path~(a) (139 instances), the multi-funded output is always a project publication.
  \item \emph{Company-patent linkage} (67): the patent is a project output in all 67 instances; the cited work is always external (\textsf{patent\_cited}).
  \item \emph{Collaboration} (318), \emph{citation cluster} (422), \emph{inventor network} (55): the output nodes are always traceable to a project; person nodes are not.
  \item \emph{Productive project} (15) and \emph{cross-stage output} (25): the project node is part of the motif definition itself.
\end{itemize}

\clearpage

\section{ARPA-E dataset context}

One reason for the poor understanding of mission-oriented R\&D programs is the difficulty of obtaining data on the outputs of these programs, and linking outputs to program-level inputs and activities. These linkages, also known as theory of change in the policy community \cite{RN1404}, is often hypothesized but rarely quantified. Fortunately, data on some challenge-led R\&D programs improved.  

Advanced Research Projects Agency - Energy (ARPA-E), established as a grant-making organization within the US department of Energy (DOE) that mimics the active project management style of the Defense Advanced Research Projects Agency (DARPA), is one of the few challenge-led R\&D agency that publicly discloses their projects’ intellectual property outputs. As one of the earliest DARPA clones, ARPA-E focuses on civilian energy innovations \cite{RN36}. The sociotechnical nature of the impacts aspired by ARPA-E is reflected by their policy of only approving programs impacting at least “one quad” (quadrillion British thermal units, roughly 1\% of US annual energy consumption) of energy system impact \cite{RN171}.

ARPA-E is now a canonical case in the mission-oriented innovation policy debate. Conceived as a civilian analogue of DARPA, it empowers program directors to pursue high-risk, high-reward energy projects using active selection, staged funding, and hands-on portfolio management \citep{Azoulay2018,GoldsteinKearney2016}. The National Academies review concluded that ARPA-E had already generated substantial scientific and technical outputs relative to its budget and that many projects attracted significant follow-on public and private finance \citep{NAS2017_ARPAE,BeyondSingleStimulus2022}. Subsequent work shows that ARPA-E-backed firms patent at higher rates than comparable cleantech start-ups, while also underscoring the difficulty of tracking long-run commercialization and spillovers \citep{Doblinger2019,StartupsARPAE2020}. Recent surveys of energy-innovation evaluation highlight the limits of prevailing metrics, which focus on publications and patents and treat projects as independent units \citep{Pless2020,IEAPless2022}. The publicly released project impact sheets therefore provide a valuable test bed for more structural, network-based evidence.

{Prior conversations with ARPA-E program directors suggest that ARPA-E-like agencies often operate around TRL3--6 \cite{RN1356}. That qualitative context motivates our interest in cross-boundary structural patterns, but the dataset used here does not directly observe program-director intent, money flows after award, or realized commercial uptake. The network therefore captures structural traces that may be consistent with those processes, not direct measurements of them.}
We aim to demonstrate network-based evidence of R\&D programs and practically improve R\&D program planning and impact assessment with a detailed analysis of all the stated outputs at ARPA-E, spanning 23 programs and 61 projects between 2010 and 2020 \cite{RN2975,RN2974,RN2973}. 

 We chose this ARPA-E dataset due to data availability and external validity. Although many new R\&D program agencies \cite{RN1361,RN2771,RN2859} acknowledge DARPA as their inspirations, detailed programmatic outputs of DARPA is rarely disclosed due to their defense nature. On the other hand, the civilian nature of ARPA-E means programmatic process \cite{RN3048,RN259} and outputs  \cite{RN2975,RN2974,RN2973} are more open to public assessments \cite{RN171}, including one at the time of writing \cite{RN3049}. Compared to the Department of Defense which accounts for over 50\% of US federal R\&D spending, the Department of Energy is transferable to a wider range of new challenge-led R\&D agencies that typically have more modest budgets \cite{RN1282,RN3065}. The ARPA-E model has also influenced other new focused research organizations. For instance, the founding CEO of UK challenge-led ARIA’s was a founding program director at ARPA-E, making an in-depth analysis of ARPA-E data more pertinent.
\clearpage

\section{Summary statistics}

\begin{table}[!ht]
  \small
  \centering 
  \caption{Summary statistics of ARPA-E program networks.}
  \label{tab:summary_stats}
\begin{threeparttable}
  \begin{tabular}{lrrrr}
    \toprule
    \textbf{Program} & \textbf{People nodes} & \textbf{Places nodes} & \textbf{Things nodes} & \textbf{Network size} \\
    \midrule
    ADEPT & 104 & 37 & 199 & 340 \\
    AMPED & 66 & 27 & 89 & 182 \\
    BEEST & 29 & 13 & 50 & 92 \\
    DELTA & 9 & 8 & 25 & 42 \\
    Electrofuels & 21 & 13 & 43 & 77 \\
    GENI & 54 & 37 & 132 & 223 \\
    GRIDS & 44 & 22 & 103 & 169 \\
    HEATS & 35 & 22 & 34 & 91 \\
    IMPACCT & 72 & 39 & 89 & 200 \\
    METALS & 31 & 10 & 72 & 113 \\
    MONITOR & 24 & 15 & 30 & 69 \\
    MOVE & 6 & 4 & 23 & 33 \\
    OPEN2009 & 76 & 36 & 201 & 313 \\
    OPEN2012 & 222 & 95 & 365 & 682 \\
    OPEN2015 & 9 & 10 & 24 & 43 \\
    PETRO & 137 & 52 & 138 & 327 \\
    RANGE & 47 & 28 & 67 & 142 \\
    REACT & 51 & 20 & 62 & 133 \\
    REBELS & 20 & 10 & 29 & 59 \\
    REMOTE & 13 & 9 & 16 & 38 \\
    Solar\_ADEPT & 23 & 18 & 43 & 84 \\
    TERRA & 51 & 33 & 93 & 177 \\
    TRANSNET & 12 & 15 & 28 & 55 \\
    \midrule
    $\Sigma$ & 1156 & 573 & 1955 & 3684 \\
    \bottomrule
  \end{tabular}
\begin{tablenotes}\footnotesize
\item {Row totals are per-program counts. Repeated people, organizations, and outputs are therefore counted once per program rather than once for the whole portfolio, which is why the row sums ($\Sigma$: 1,156 people; 573 places; 1,955 things) exceed the deduplicated denominators used in the overlap analysis. In the main text, person-level overlap is reported against 1,056 researcher nodes and organization-level overlap against 324 organization nodes computed from the combined portfolio graph. The \textbf{Things} column includes each program's stated paper and patent outputs together with their first-degree forward and backward citations, so it is not comparable to the 151 direct outputs disclosed in the impact sheets. Although our sample captures all information from ARPA-E's published program impact sheets \cite{RN2975,RN2973,RN2974}, it represents only a disclosed subset of ARPA-E's total activity.}
\end{tablenotes}
\end{threeparttable}
\end{table}
\clearpage

\section{{Detector overlap matrix}}
\label{app:detector_overlap}

\begin{table}[!ht]
\small
\centering
\caption{{Instance-level detector overlap. Cell $(A,B)$ gives the number of type-$A$ motif instances that share at least one output node (publication, patent, or project) with any type-$B$ instance. Diagonal entries are total counts per type. The heaviest off-diagonal overlaps are between multi-funder linkage (MFLink) and co-funding linkage (CoFLink), which share an underlying co-funded output in 567 of 570 MFLink instances.}}
\label{tab:detector_overlap}
\begin{tabular}{l*{8}{r}}
\toprule
 & \textbf{MFLink} & \textbf{CoFLink} & \textbf{CPLink} & \textbf{CSOut} & \textbf{Collab} & \textbf{CitClust} & \textbf{ProdProj} & \textbf{InvNet} \\
\midrule
MFLink   & 570 & 567 &   0 &  48 & 557 & 333 &  85 &   0 \\
CoFLink  & 161 & 170 &   0 &   7 & 161 & 120 &  15 &   0 \\
CPLink   &   0 &   0 &  67 &  23 &  52 &   0 &  19 &  46 \\
CSOut    &   9 &   7 &  23 &  25 &  21 &   7 &  15 &  13 \\
Collab   & 137 & 151 &  52 &  31 & 318 & 159 &  45 &  42 \\
CitClust & 366 & 365 &   0 &  36 & 415 & 422 &  47 &   0 \\
ProdProj &   9 &   8 &  14 &  15 &  15 &   8 &  15 &   6 \\
InvNet   &   0 &   0 &  49 &  37 &  50 &   0 &  20 &  55 \\
\bottomrule
\end{tabular}
\end{table}
\clearpage

\section{{Exploratory comparison of OPEN calls and targeted programs}}
\label{app:open_vs_targeted}

Table~\ref{tab:open_vs_targeted} reports normalized motif shares (percentage of each program's total motif count) for all 23 programs, grouped by call type. The bottom rows give group means and the results of Mann-Whitney $U$ tests and two-sided permutation tests (10,000 draws) comparing the two groups on each motif type. With $n=3$ OPEN calls, all tests are underpowered; no motif type reaches $p < 0.05$.

\begin{table}[!ht]
\small
\centering
\caption{Normalized motif shares (\% of program total) by call type, with group comparisons. \textbf{O} = OPEN call, \textbf{T} = Targeted. programs with fewer than 5 total motifs (DELTA, MOVE, REMOTE) are included for completeness but their percentages are unreliable. $\dagger$ = $p < 0.10$; n.s. = not significant ($p \geq 0.10$) on both Mann-Whitney and permutation tests.}
\label{tab:open_vs_targeted}
\begin{threeparttable}
\begin{tabular}{lrrrrrrrrrr}
\toprule
\textbf{program} & \textbf{Type} & \textbf{$N$} & \textbf{Total} & \textbf{Collab} & \textbf{CitClust} & \textbf{InvNet} & \textbf{MFLink} & \textbf{CoFLink} & \textbf{CPLink} & \textbf{CSOut} \\
\midrule
ADEPT        & T & 332 & 73  & 32 &  1 &  4 & 38 & 11 &  5 &  4 \\
AMPED        & T & 174 & 79  & 28 & 38 &  0 & 22 &  5 &  3 &  3 \\
BEEST        & T &  88 & 39  & 21 &  5 & 21 & 26 & 13 & 10 &  3 \\
DELTA        & T &  38 &  2  &100 &  0 &  0 &  0 &  0 &  0 &  0 \\
Electrofuels & T &  73 & 28  & 18 & 11 &  7 & 39 & 11 &  7 &  4 \\
GENI         & T & 213 & 51  & 33 & 10 & 10 & 25 & 10 &  8 &  2 \\
GRIDS        & T & 159 & 52  & 25 & 19 &  6 & 27 &  8 & 12 &  4 \\
HEATS        & T &  87 & 48  & 19 & 17 &  0 & 56 &  8 &  0 &  0 \\
IMPACCT      & T & 194 & 84  & 18 &  8 &  0 & 56 & 11 &  2 &  2 \\
METALS       & T & 107 & 29  & 31 & 45 &  0 & 10 &  3 &  3 &  3 \\
MONITOR      & T &  65 & 27  & 22 &  7 &  0 & 56 & 15 &  0 &  0 \\
MOVE         & T &  29 &  2  &  0 &  0 &  0 &  0 &  0 & 50 & 50 \\
PETRO        & T & 317 & 147 & 14 &  2 &  0 & 70 & 12 &  1 &  1 \\
RANGE        & T & 136 & 101 & 18 & 32 &  0 & 41 & 10 &  0 &  0 \\
REACT        & T & 127 &  98 & 17 & 33 &  0 & 38 & 12 &  0 &  0 \\
REBELS       & T &  55 &  24 & 25 & 12 &  0 & 50 & 12 &  0 &  0 \\
REMOTE       & T &  34 &   3 &100 &  0 &  0 &  0 &  0 &  0 &  0 \\
Solar\_ADEPT  & T &  80 &  81 & 14 & 48 &  0 & 16 & 22 &  0 &  0 \\
TERRA        & T & 165 &  32 & 28 &  3 &  0 & 59 &  9 &  0 &  0 \\
TRANSNET     & T &  51 &  14 & 21 &  0 &  0 & 57 & 21 &  0 &  0 \\
\midrule
OPEN2009     & O & 295 & 116 & 25 &  9 & 23 &  8 &  3 & 28 &  5 \\
OPEN2012     & O & 658 & 498 & 14 & 44 &  1 & 27 & 10 &  2 &  1 \\
OPEN2015     & O &  39 &  14 & 14 & 21 &  0 & 50 & 14 &  0 &  0 \\
\midrule
\textit{OPEN mean}     & & & & 17.8 & 24.6 & 8.2 & 28.4 & 9.1 & 9.7 & 2.0 \\
\textit{Targeted mean} & & & & 29.2 & 14.6 & 2.4 & 34.3 & 9.7 & 5.1 & 3.8 \\
\textit{Difference (pp)} & & & & $-11.4$ & $+10.0$ & $+5.9$ & $-6.0$ & $-0.6$ & $+4.7$ & $-1.8$ \\
\textit{MW $p$-value}    & & & & 0.235 & 0.253 & 0.179 & 0.715 & 1.000 & 0.663 & 0.663 \\
\textit{Perm $p$-value}  & & & & 0.355 & 0.277 & 0.078$\dagger$ & 0.626 & 0.885 & 0.425 & 0.710 \\
\textit{Significance}    & & & & n.s. & n.s. & $\dagger$ & n.s. & n.s. & n.s. & n.s. \\
\bottomrule
\end{tabular}
\begin{tablenotes}\footnotesize
\item Shares are computed as each motif type's count as a percentage of the program's total motifs. Productive project is omitted from the table to save space (shares $<2\%$ for all programs; group difference $-0.9$~pp, $p = 0.57$). Tests use normalized shares; Mann-Whitney $U$ (two-sided) and permutation test (10,000 draws, seed 42). With $n=3$ OPEN calls, power is very low and no result should be treated as confirmatory.
\end{tablenotes}
\end{threeparttable}
\end{table}

\clearpage

\section{Internal program architecture and capacity assembly}

\subsection{People}

To ground the abstract discussion of R\&D networks, we make repeated use
of the AMPED network as reference, with the 20 other networks included in the supplementary information. Fig. \ref{fig:amped_people} plots the people directly involved in the AMPED program and label them by their relation with the program director. We were able to trace the program director’s (Ilan Gur) recruitment of three performers (Regan Zane, Ajay Raghavan, Aaron Knobloch), who in turn brought in their own 58 collaborators when publishing or patenting. We further applied social network metrics to the largest connected component (16 people) of the AMPED people network
showing that Ajay Raghavan (PI) and co-authors Alexander Lochbaum and Peter Kiesel achieve betweenness, closeness, and eigenvector centralities of 1.0 within this connected component. In this local graph they are the most central nodes linking otherwise disjoint research teams, although the metric does not by itself demonstrate wider knowledge brokerage beyond the observed network. The median betweenness centrality within this component is 0.11, indicating a concentrated centrality structure in which most participants occupy more peripheral positions.

{Using this people network, we can examine how one disclosed program network relates to grounded theories about R\&D program management (Table \ref{tab:folk_knowledge}). One hypothesis by Fuchs \cite{RN445} is that program directors take an activist role in architecting and governing the social network of innovation. While our network snapshot focuses on program performers, the concentration of centrality among a few leads is at least consistent with the idea that ARPA-E staff seed networks that rely on explicitly recruited connectors. Another descriptive observation is that none of the 62 people across the three projects within the same program overlap, which is consistent with performer diversification within this single program snapshot.}

\begin{figure}[h!]
  \centering
  \includegraphics[width=0.9\textwidth]{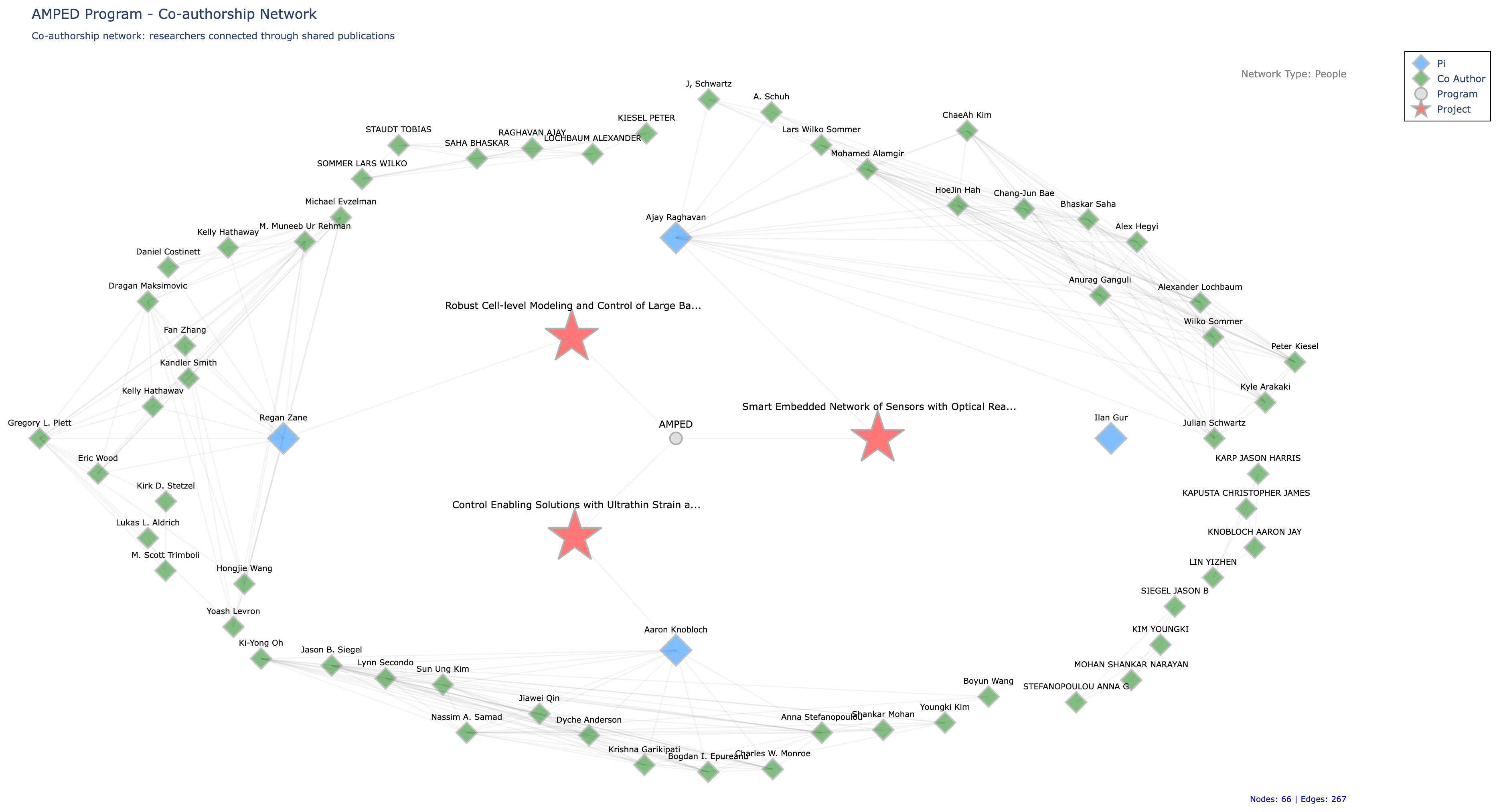}
  \caption{People network of AMPED. Nodes are the program director, project PI, and co-authors of PIs. Edges are the observed funding and collaboration relations. An interactive version of this figure is available at \texttt{Fig1a\_AMPED\_people.html}.}
  \label{fig:amped_people}
\end{figure}

\subsection{Institutions}

\begin{figure}[h!]
  \centering
  \includegraphics[width=0.9\textwidth]{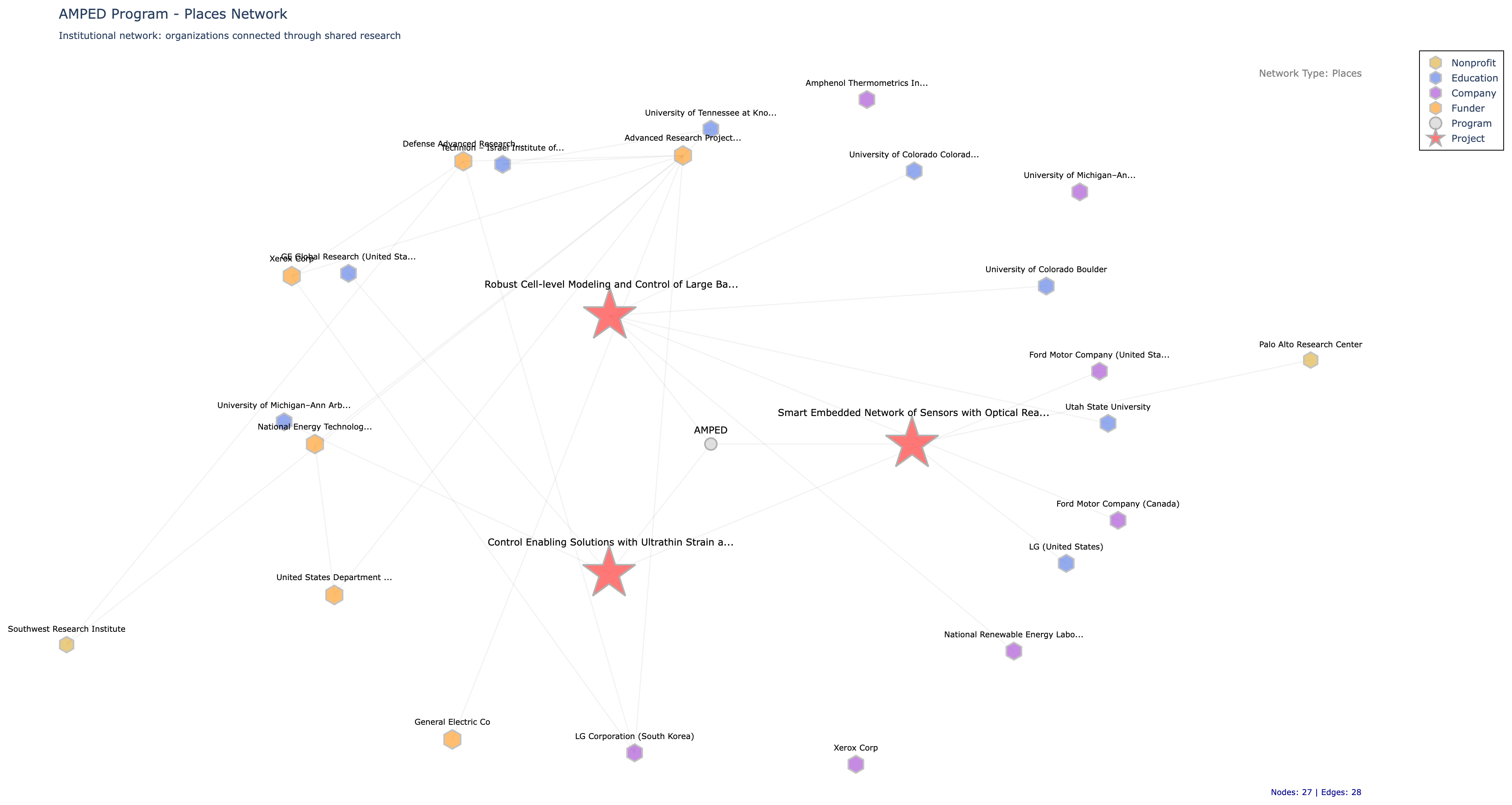}
  \caption{Places network of AMPED program. Nodes are the host institutions of ARPA-E program PIs and funders that co-fund the PIs. These institutions are labelled by their organization type recorded on the Research Organization Registry. Edges are the observed funding and co-funding relations. An interactive version of this figure is available at \texttt{Fig1b\_AMPED\_places.html}.}
  \label{fig:amped_places}
\end{figure}

{Moving onto the places network, Fig. \ref{fig:amped_places} plots the host institutions and funders of AMPED’s performers and the performers’ collaborators. The nodes are labelled by the type of organization as classified by the \texttt{Research Organization Registry}. In this disclosed network, ARPA-E is connected to 8 research and academic institutions, 7 companies, 6 funding agencies (including DARPA), and 2 nonprofit research organizations through observed funding and co-funding relations. For instance, Xerox is connected with DARPA and LG via co-funding of research projects. This is best read as a shared funding context rather than proof that ARPA-E itself directly mobilized Xerox or LG as collaborators. These network motifs are explored in greater detail in Section \ref{ch:results_motif} and can be quantified at scale: across AMPED’s three projects the mean Shannon diversity of organization types is 0.67, with universities comprising 59\% of participating institutions and companies 24\%. These diversity metrics provide a replicable baseline for granular benchmarking of public-private-university partnerships across R\&D programs.}

{It has been hypothesized that challenge-led R\&D programs require coordinated investment across disciplines, institutions, and infrastructure to scale up new technologies. One such framework is the “island-bridge” model \cite{RN3069,RN36}, whereby organizations such as ARPA-E would organize as a largely autonomous “island” to experiment while extending many “bridges” to leaderships and other agencies. Across all 23 program networks, ARPA-E is connected to 73 private companies, 45 other government agencies, 24 national labs, and 2 philanthropies. These counts document the breadth of the observed institutional neighborhood, not the causal effect of ARPA-E on those organizations.}

\subsection{Hand-off pathways and propagation}

\begin{figure}[h!]
  \centering
  \includegraphics[width=0.9\textwidth]{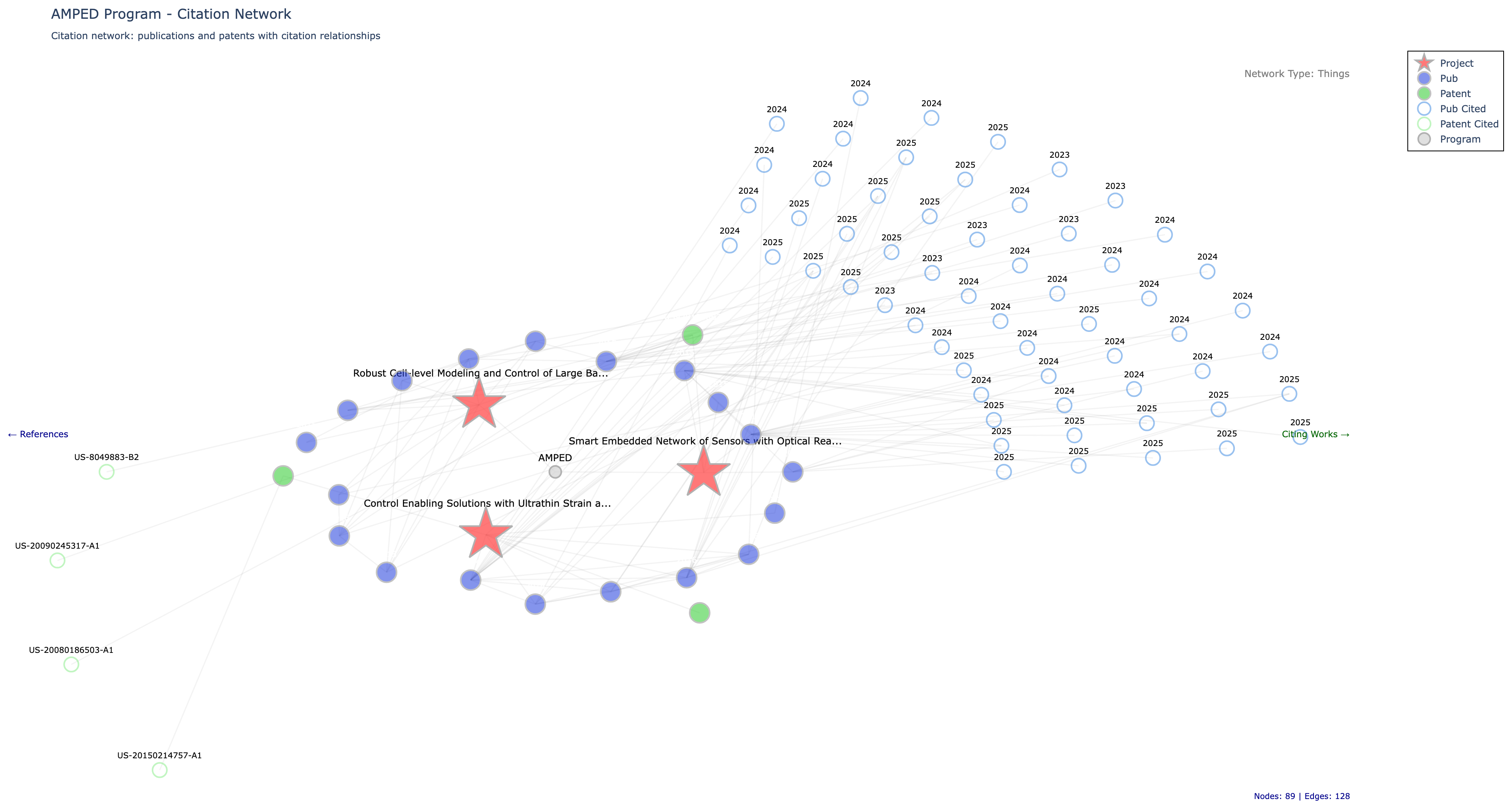}
  \caption{Citation (things) network of AMPED program. Nodes on the far left are documents cited by ARPA-E outputs; nodes on the far right are documents that cite ARPA-E outputs. Edges are the funding and citation relations. An interactive version of this figure is available at \texttt{Fig1c\_AMPED\_things.html}.
  \label{fig:amped_things}}
\end{figure}

{To map how ARPA-E programs interact with existing knowledge and produce new knowledge, Fig. \ref{fig:amped_things} shows the immediate publication and patent outputs from the AMPED program, enriched with forward and backward citations. The possibility to tailor R\&D programs' ``things'' networks is broad and depends on the specific investment or policy question one has.}

{An example analysis that networks are well-placed to perform is the program-specific geographical reach of citation-linked knowledge. In the case of the AMPED program, \$4M of project funding produced 4 publications that cited 23 publications produced in the US, China, Korea, Switzerland, Austria, and Germany. At the time of writing, these 4 publications are in turn cited by publications and patents in the US, China, and Singapore. Even though ARPA-E only funded domestically, the citation network shows that AMPED outputs sit within a cross-border knowledge neighbourhood.}

Should one be interested in ARPA-E's citation-linked domain reach, the citation networks of the 65 publications produced by ARPA-E's 23 programs mostly draw on knowledge in materials engineering (56\% of backward citations; 287 cited papers). Interestingly, most first-degree papers that cite ARPA-E's largely materials-engineering research are related to physical chemistry (52\% of forward citations; 687 citing publications). Future studies can use things networks to describe disciplinary positioning and reach of R\&D agencies more systematically.

\begin{figure}[h!]
  \centering
  \includegraphics[width=0.9\textwidth]{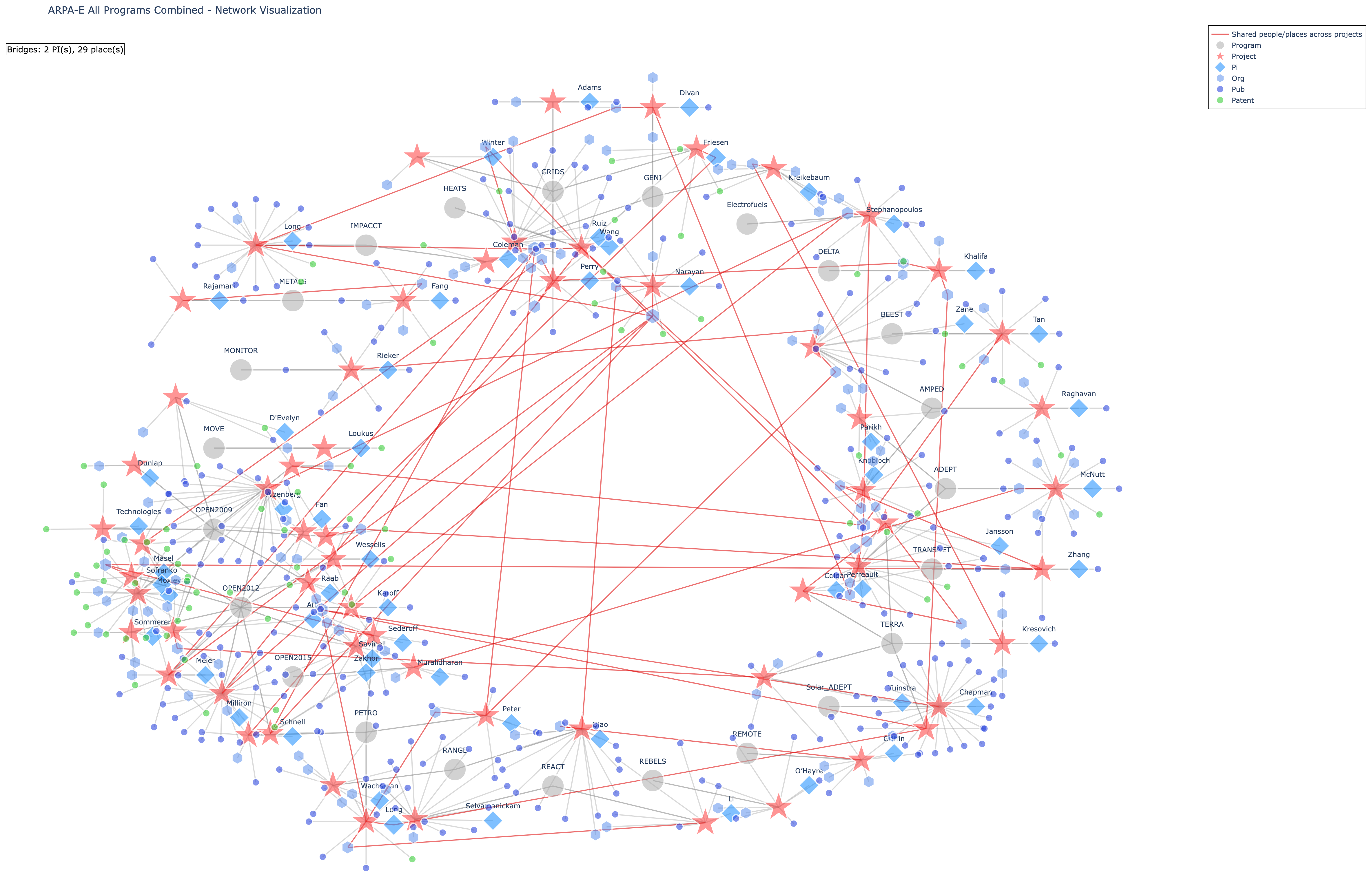}
  \caption{{Complete network of ARPA-E people, places, and things. Red lines highlight edges involving shared people or places that bridge across two or more ARPA-E programs in this plotted all-program network (2~PIs and 29~institutions in this visualization). This figure-specific bridge count is narrower than the broader overlap totals reported in the main text, which are computed from the combined portfolio graph using deduplicated node categories and include recurrent co-funder nodes. Grey edges are intra-program connections. Node types follow the legend: grey circles = programs, red stars = projects, blue diamonds = PIs, light blue circles = organizations, green = publications, and small circles = patents and cited works. Data is based solely on project outputs detailed in \cite{RN2975,RN2974,RN2973}. An interactive version is available at \texttt{Fig2b\_all\_programs\_combined.html}.}}
  \label{fig:arpae_all}
\end{figure}

\section{Robustness}
\label{sec:robustness}

To ensure that the observed motif prevalences reflect genuine structural signals rather than artefacts of measurement design, we conduct three robustness analyses.

\paragraph{{Null-model contrast.}}
{For each program, we generate 100 randomized graphs using a typed, degree-preserving edge-swap algorithm that preserves (i)~the number and type of nodes, (ii)~the in-degree and out-degree of every node within each edge type, and (iii)~the node-type constraints on edge endpoints. Each randomized graph receives the same DAG-enforcement preprocessing (Step~2 of the detection pipeline) before motif counting, ensuring that any bias introduced by citation-cycle removal is identical in the observed and random graphs. We then recount all eight motifs in each random graph and compute $z$-scores: $z = (\text{observed} - \mu_{\mathrm{rand}}) / \sigma_{\mathrm{rand}}$. We report $z$-scores as standardized effect sizes for cross-program comparability. A p-value would require either an additional distributional assumption or a direct Monte Carlo tail calculation; with only 100 randomizations, those empirical tail probabilities would also be coarse. Because motif-count distributions are often non-Gaussian, discrete, or nearly degenerate, and because 184 comparisons are made across 23~programs and 8 motif types, these $z$-scores are used descriptively and should not be interpreted as formal p-values or significance tests.}

{Four} important caveats apply to the null-model comparison.

{First, \emph{temporal admissibility}. Citations are temporally constrained: a document published in year $t$ can only cite documents published before $t$. The current null randomizes citation endpoints among same-category nodes without enforcing this temporal ordering, then removes any resulting cycles via DAG enforcement. This means the null can generate temporally impossible citation edges (e.g.\ a 2012 publication citing a 2018 publication) which are then partly pruned. A cleaner baseline would restrict each citation swap to pairs of target nodes whose publication dates are both earlier than the citing node's date, preserving temporal admissibility by construction rather than by post-hoc pruning. The citation-cluster $z$-scores reported below should therefore be interpreted as contrasts against a degree-preserving null that is not fully time-respecting, which may overstate or understate the true elevation relative to a temporally constrained baseline. Implementing such a time-respecting null and increasing the number of randomizations beyond 100 are priorities for future work.}

\begin{figure}[h!]
  \centering
  \includegraphics[width=\textwidth]{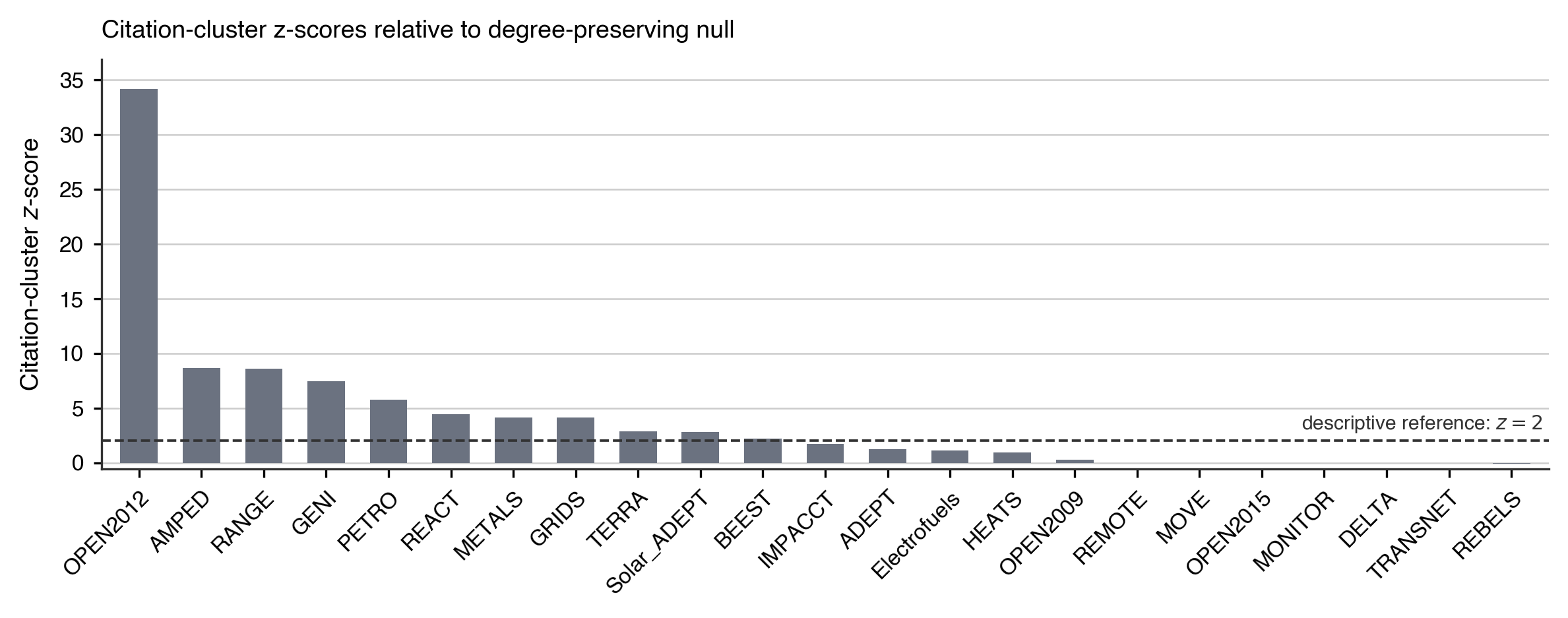}
  \caption{{{\textbf{Citation-cluster $z$-scores relative to a degree-preserving null model} ($n=100$ randomizations per program). Bars show the standardized null contrast for each program's citation-cluster count relative to the randomized baseline. The dashed line at $z=2$ is a descriptive reference benchmark only; it is not a p-value threshold or formal significance cutoff.}}}
  \label{fig:zscore_heatmap}
\end{figure}

Second, for degree-threshold motifs: \emph{collaboration} (outputs with $\geq 3$ incoming \textsf{authored}/\textsf{invented} edges), \emph{inventor network} (inventors with $\geq 2$ outgoing \textsf{invented} edges), and path~(a) of \emph{co-funding linkage} (outputs with $\geq 2$ incoming \textsf{funds} edges from distinct funders): a per-node, per-edge-type degree-preserving null renders counts invariant by construction: each motif is a deterministic function of the preserved degrees, so $\sigma_{\mathrm{rand}} \approx 0$ and $z \approx 0$ by design. Their near-zero $z$-scores cannot be used to infer degree-structure causation; they are trivially invariant under the null and are excluded from Fig.~\ref{fig:zscore_heatmap}, which focuses on the citation-cluster signal. The null comparison is most informative for motifs that depend on higher-order wiring, such as citation clusters and cross-boundary funding patterns. If non-zero $\sigma_{\mathrm{rand}}$ is observed for these degree-threshold motifs, it would indicate that the null is not preserving per-node, per-edge-type degrees as intended.

{Third, \emph{number of randomizations}. With $n=100$ randomizations per program, Monte Carlo tail probabilities are coarse (resolution of 0.01) and empirical standard deviations may not have converged for motif types with heavy-tailed distributions. The $z$-scores are therefore best treated as order-of-magnitude contrasts rather than precise effect sizes.}

{Fourth, \emph{DAG enforcement bias}. The cycle-removal step deletes citation edges from strongly connected components. Because the null model can create more cycles than the observed graph (which is already largely time-ordered), the null graphs may systematically lose more citation edges during DAG enforcement, which could deflate null-model citation-cluster counts and inflate the observed $z$-scores. The magnitude of this bias is unknown without a time-respecting null that avoids post-hoc edge deletion entirely.}

{{Fig.~\ref{fig:zscore_heatmap} shows citation-cluster $z$-scores across all 23 programs. For all other motif types, counts reproduce at near-expected levels under randomization (mean $|z| < 1$), consistent with those patterns being driven primarily by degree structure, i.e., by which actors and outputs are present, rather than by higher-order wiring. Citation clusters are the principal exception. Eleven programs lie above the dashed $z=2$ reference line, but that line is shown only as a visual benchmark for unusually large positive null contrasts, not as a p-value cutoff.}}

{Among the communication-labelled motifs, multi-funder linkages show relatively high $z$-scores in HEATS ($z=2.4$) and low values in MONITOR ($z=-2.6$). OPEN2012 co-funding linkage is strongly below its null expectation ($z=-4.9$), meaning that OPEN2012's large multi-funder graph would predict even more such instances than observed; OPEN2009 cross-stage output is likewise below its null expectation ($z=-10.0$). As a further structural check, the near-equal communication/configuration split (50.7\%/49.3\%) is consistent with the null model. Because most motif types are broadly consistent with degree-preserving randomization, that balance should be interpreted as a feature of the network's typed degree distribution rather than as evidence of any higher-order structural preference.}

\paragraph{Citation-cluster enumeration.}
The motif detection pipeline enumerates all triangles in the undirected citation projection without any per-program cap. Six programs produce more than 10 citation-cluster triangles (OPEN2012: 218, Solar\_ADEPT: 39, RANGE: 32, REACT: 32, AMPED: 30, METALS: 13), contributing 422 citation-cluster instances in total. This full enumeration ensures that reported counts are exact rather than conservative lower bounds.

\paragraph{Program fingerprint clustering.}
To test whether pattern signatures correspond broadly to known program characteristics, we cluster programs by centered log-ratio (CLR) transformed motif-share vectors using Euclidean distance with Ward linkage, and project them into two dimensions via PCA (Fig.~\ref{fig:clustering}). Color-coded thematic categories (extracted from ARPA-E project impact sheets) are overlaid on both the dendrogram and the scatter plot.

{Because the CLR transform is undefined for exact zeros and 63 of 184 cells (34.2\%) in the $23 \times 8$ motif-count matrix are zero, we apply additive smoothing with a pseudocount of $\delta = 0.5$ before computing shares: $x'_{ij} = (x_{ij} + \delta) / \sum_k (x_{ik} + \delta)$. To verify that results are not driven by this choice, we repeated the Ward clustering at pseudocounts of 0.1, 0.25, 1.0, and 2.0. The broad cluster assignments are stable across the range $\delta \in [0.5, 2.0]$: the same core groups (citation-dense programs such as OPEN2012, RANGE, REACT, Solar ADEPT; and configuration-heavy programs such as METALS, GENI, GRIDS) emerge at each setting. At $\delta < 0.25$, the three very small programs (DELTA, MOVE, REMOTE) migrate between clusters, consistent with their sparse count vectors being sensitive to smoothing. We therefore treat $\delta = 0.5$ as a reasonable default and flag the small-program instability.}

{Several interpretable groupings emerge among programs with sufficient motif counts.} Communication-heavy programs (OPEN2012, RANGE, REACT, AMPED, Solar ADEPT, PETRO, TERRA) cluster together at low Ward distance, while configuration-heavy programs (METALS, OPEN2009, GENI, GRIDS, ADEPT, BEEST) form a separate cluster. Small programs with sparse pattern vectors (DELTA, REMOTE, MOVE, OPEN2015) form a distinct group at the bottom of the dendrogram; {this cluster is driven by sparsity rather than by a substantive structural commonality and should not be interpreted comparatively.} The broad alignment between pattern-derived clusters and known program themes {among the remaining programs} is best read as descriptive convergence or face validity, not as strict external validation.

\begin{figure}[h!]
  \centering
  \includegraphics[width=\textwidth]{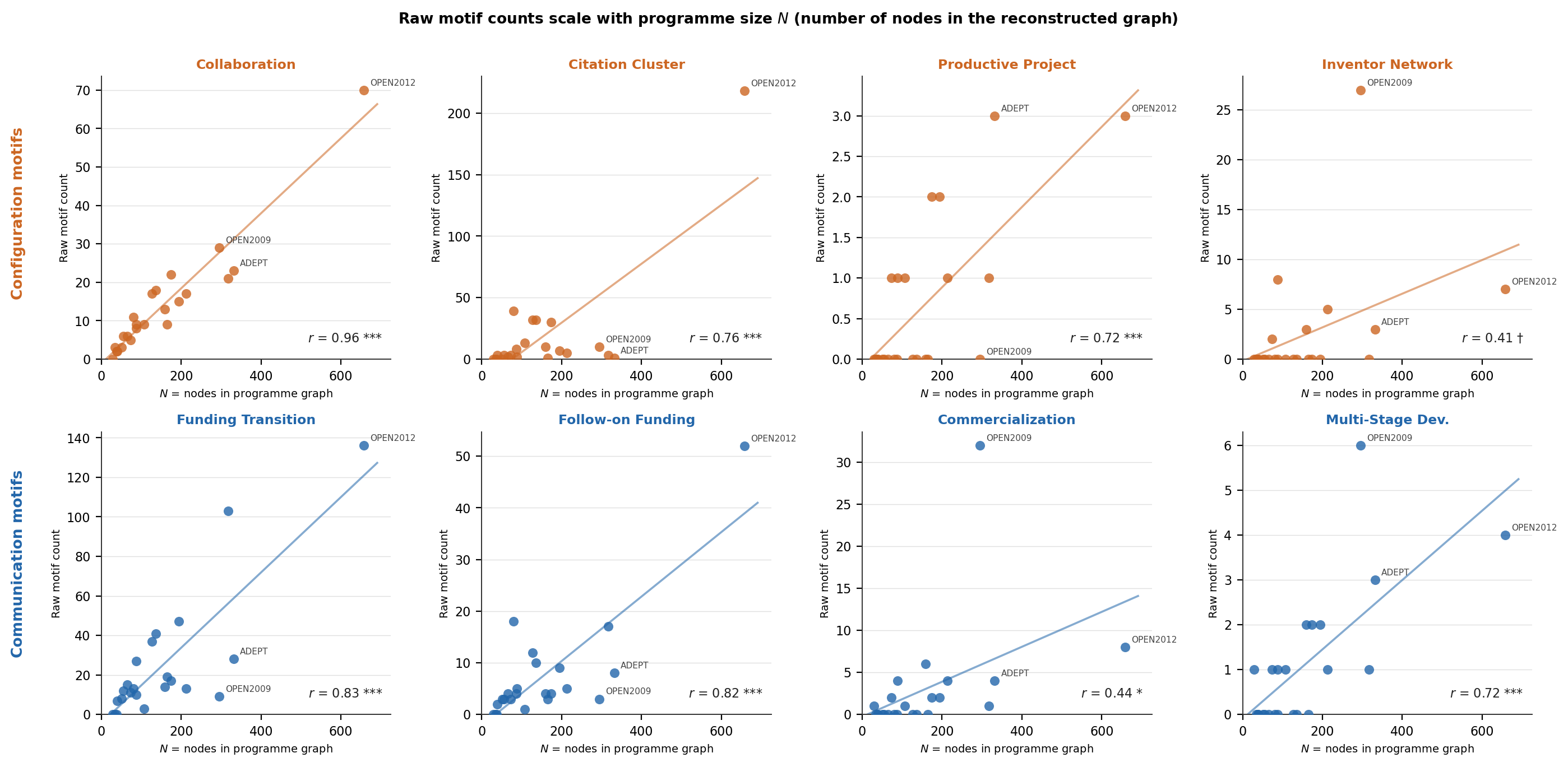}
  \caption{{\textbf{Raw pattern counts scale with program size $N$, but normalized shares do not.} Each panel shows raw pattern count versus $N$ (number of nodes in the program's reconstructed graph) across all 23 programs ($n=23$), with Pearson $r$ shown for descriptive context. $N$ is defined as the total number of typed nodes (persons, organizations, outputs, projects) in the program-induced subgraph after DAG enforcement. The figure is used to show size dependence in raw counts and the relative stability of normalized shares; it is not intended as a stand-alone inferential result.}}
  \label{fig:motif_vs_N}
\end{figure}

A noteworthy property of the normalized fingerprints is their \emph{scale-independence}. Program size (number of nodes in the reconstructed graph) is a strong predictor of total raw motif count across the 23 programs (Pearson $r = 0.90$, $n = 23$), confirming that larger programs generate proportionally more structural events. However, program size has near-zero correlation with each program's normalized motif \emph{shares} (mean $|r| = 0.13$ across all eight types, maximum $|r| = 0.24$ for citation clusters), indicating that relative fingerprint composition is independent of program size. A large program is not structurally similar to a small one simply by virtue of being large; motif composition is instead determined by design choices, technology domain, and performer mix. This scale-independence is a methodologically desirable property: the normalized fingerprint is a valid comparative diagnostic across programs spanning more than a twenty-fold range in size (33 to 682 nodes), and the PCA clustering is not an artefact of size differences between programs. Fig.~\ref{fig:motif_vs_N} shows the per-motif scatter plots.

\end{document}